\pdfoutput=1
\documentclass[usenatbib]{mn2e}
\usepackage{apjfonts}
\usepackage{amssymb}
\usepackage{amsmath}
\usepackage{ctable}
\usepackage{url}
\usepackage{fixltx2e} 

\newcommand{\be}{\begin{equation}}
\newcommand{\ee}{\end{equation}}

\newcommand{\SM}{SM12}
\newcommand{\DV}{\Delta V}
\newcommand{\DVa}{\Delta \tilde{V}}
\newcommand{\Sentropy}{S02}
\newcommand{\NNb}{N_{\rm NGB}}
\newcommand{\wengenurl}{\url{http://ascl.net/1305.006}}

\newcommand\plotone[1]
 {\centering \leavevmode \includegraphics[width={0.99\columnwidth}]{#1}}

\newcommand\plotonesize[2]
 {\centering \leavevmode \includegraphics[width={#2\columnwidth}]{#1}}
\newcommand{\plotsidesize}[2]
 {\centering \leavevmode \includegraphics[width={#2\textwidth}]{#1}}
\newcommand{\acknowledgments}{\begin{small}\section*{Acknowledgments}\end{small}}
\newcommand\altaffilmark[1]{$^{#1}$}
\newcommand\altaffiltext[1]{$^{#1}$}
\title[Lagrangian SPH \&\ Mixing]{A General Class of Lagrangian Smoothed Particle Hydrodynamics Methods and Implications for Fluid Mixing Problems
\vspace{-0.5cm}}

\author[Hopkins et al.]{
\parbox[t]{\textwidth}{ 
Philip F. Hopkins\altaffilmark{1}\thanks{E-mail:phopkins@astro.berkeley.edu}
} 
\vspace*{6pt} \\
\altaffiltext{1}{Department of Astronomy, University of California
  Berkeley, Berkeley, CA 94720\vspace{-1.1cm}} \\
}

\date{Submitted to MNRAS, June, 2012\vspace{-0.6cm}}
\begin{document}
\maketitle
\label{firstpage}

\begin{abstract}

Various formulations of smooth-particle hydrodynamics (SPH) have been proposed, intended to resolve certain difficulties in the treatment of fluid mixing instabilities. Most have involved changes to the algorithm which either introduce artificial correction terms or violate what is arguably the greatest advantage of SPH over other methods: manifest conservation of energy, entropy, momentum, and angular momentum. Here, we show how a class of alternative SPH equations of motion (EOM) can be derived self-consistently from a discrete particle Lagrangian -- guaranteeing manifest conservation -- in a manner which tremendously improves treatment of these instabilities and contact discontinuities. Saitoh \&\ Makino recently noted that the volume element used to discretize the EOM does not need to explicitly invoke the mass density (as in the ``standard'' approach); we show how this insight, and the resulting degree of freedom, can be incorporated into the rigorous Lagrangian formulation that retains ideal conservation properties and includes the ``$\nabla h$'' terms that account for variable smoothing lengths. We derive a general EOM for any choice of volume element (particle ``weights'') and method of determining smoothing lengths. We then specify this to a ``pressure-entropy formulation'' which resolves problems in the traditional treatment of fluid interfaces. Implementing this in a new version of the {\small GADGET} code, we show it leads to good performance in mixing experiments (e.g. Kelvin-Helmholtz \&\ ``blob'' tests). And conservation is maintained even in strong shock/blastwave tests, where formulations without manifest conservation produce large errors. This also improves the treatment of sub-sonic turbulence, and lessens the need for large kernel particle numbers. The code changes are trivial and entail no additional numerical expense. This provides a general framework for self-consistent derivation of different ``flavors'' of SPH.

\end{abstract}

\begin{keywords}
methods: numerical --- hydrodynamics --- instabilities --- turbulence --- cosmology: theory
\vspace{-1.0cm}
\end{keywords}

\vspace{-1.1cm}
\section{Introduction}
\label{sec:intro}

Smoothed particle hydrodynamics (SPH) is a method for solving the equations of hydrodynamics \citep[in which Lagrangian discretized mass elements are followed;][]{lucy:1977.sph,gingold.monaghan:1977.sph} which has found widespread application in astrophysical simulations and a range of other fields as well \citep[for recent reviews, see][]{rosswog:2009.sph.review,springel:2010.sph.review,price:2012.sph.review}. 

The popularity of SPH owes to a number of properties: compared to many other methods, it is numerically very robust (stable), trivially allows the tracing of individual fluid elements (Lagrangian), automatically produces improved resolution in high-density regions without the need for any ad-hoc pre-specified ``refinement'' criteria (inherently adaptive), is Galilean-invariant, couples properly and conservatively to N-body gravity schemes, exactly solves the particle continuity equation,\footnote{This is the continuity equation for a {\em discretized} particle field. Exactly solving the continuity equation for a continuous fluid, of course, requires infinite resolution or infinite ability to distort the Lagrangian particle ``shape.''} and has excellent conservation properties. The latter character stems from the fact that -- unlike Eulerian grid methods -- the SPH equations of motion (EOM) can be rigorously and exactly derived from a discretized particle Lagrangian, in a manner that guarantees manifest and simultaneous conservation of energy, entropy, linear momentum, and angular momentum \citep[][henceforth \Sentropy]{springel:entropy}.

However, there has been considerable discussion in the literature regarding the accuracy with which the most common SPH algorithms capture certain fluid mixing processes \cite[particularly the Kelvin-Helmholtz instability; see e.g.][]{morris:1996.sph.stability,dilts:1999.sph.stability,ritchie.thomas:2001.egy.wtd.sph,marri:2003.mod.sph.cosmo.sims,okamoto:2003.shear.sph.flows,agertz:2007.sph.grid.mixing}. Comparison between SPH and Eulerian (grid) methods shows that while agreement is quite good for super-sonic flows, strong shock problems, and regimes with external forcing (e.g.\ gravity); ``standard'' SPH appears to suppress mixing in sub-sonic, thermal pressure-dominated regimes associated with contact discontinuities \citep{kitsionas:2009.grid.sph.compare.turbulence,price:2010.grid.sph.compare.turbulence,bauer:2011.sph.vs.arepo.shocks,sijacki:2011.gadget.arepo.hydro.tests}.\footnote{In fairness, we should emphasize that it has long been well-known that Eulerian grid codes, on the other hand, err on the side of {\em over}-mixing (especially when resolution is limited), and in fact this problem actually motivated some of the SPH work discussed above. This may, however, be remedied in moving-mesh approaches (though further study is needed; see e.g.\ \citealt{springel:arepo}).} The reason is, in part, that in standard SPH the kernel-smoothed density enters the EOM, and so behaves incorrectly near contact discontinuities (introducing an artificial ``surface tension''-like term) where the density is not differentiable. 

A variety of ``flavors'' (alternative formulations of the EOM or kernel estimators) of SPH have been proposed which remedy this \citep[see above and][]{monaghan:1997.sph.drag.viscosities,ritchie.thomas:2001.egy.wtd.sph,price:2008.sph.contact.discontinuities,wadsley:2008.sph.mixing.cosmology,read:2010.sph.mixing.optimization,read:2012.sph.w.dissipation.switches,abel:2011.sph.pressure.gradient.est,garciasenz:2012.integral.sph}. These approaches share an essential common principle, namely recognizing that the pressure at contact discontinuities must be single-valued (effectively removing the surface tension term). Some of these show great promise. However, many (though not all) of these formulations either introduce additional (potentially unphysical) dissipation terms and/or explicitly violate the manifest conservation and continuity solutions described above -- perhaps the greatest advantages of SPH. This can lead to severe errors in problems with strong shocks or high-Mach number flows, limited resolution, or much larger gradients between phase boundaries (J.\ Read, private communication; see also the discussion in \citealt{price:2012.sph.review,read:2012.sph.w.dissipation.switches,abel:2011.sph.pressure.gradient.est}). All of these regimes are inevitable in most astrophysically interesting problems. 

Recently however, \citet{saitoh:2012.dens.indep.sph} (henceforth \SM) pointed out that the essential results of most of these flavors can be derived self-consistently in a manner that does properly conserve energy. The key insight is that the ``problematic'' inclusion of the density in the EOM (as opposed to some continuous property near contact discontinuities) arises because of the ultimately arbitrary choice of how to discretize the SPH {\em volume} element (typically chosen to be $\sim m_{i}/\rho_{i}$). Beginning with an alterative choice of volume element, one can in fact consistently derive a conservative EOM. They propose a specific form of the volume element involving internal energy and pressure, and show that this eliminates the surface tension term and resolves many problems of mixing near contact discontinuities. 

In this paper, we develop this approach to provide a rigorous, conservative, Lagrangian basis for the formulation of alternative ``flavors'' of SPH, and show that this can robustly resolve certain issues in mixing. Although the EOM derived in \SM\ conserves energy, it was derived from an ad-hoc discretization of the hydrodynamic equations, not the discrete particle Lagrangian. As such it cannot guarantee {\em simultaneous} conservation of energy and entropy (as well as momentum and angular momentum). And the EOM they derive is conservative only for constant SPH smoothing lengths (in time and space); to allow for adaptive smoothing (another major motivation for SPH), it is necessary to derive the ``$\nabla h$'' terms which account for their variations. This links the volume elements used for smoothing in a manner that necessitates a Lagrangian derivation. And their derivation depends on explicitly evolving the particle internal energy; there are a number of advantages to adopting entropy-based formulations of the SPH equations instead. 

We show here that -- allowing for a different initial choice of which thermodynamic volume variable is discretized -- an entire extensible class of SPH algorithms can be derived from the discrete particle Lagrangian, and write a general EOM for these methods (Eq.~\ref{eqn:eom}, our key result). We derive specific ``pressure-energy'' (Eq.~\ref{eqn:eom.pressure.energy.betterh}) and ``pressure-entropy'' (Eq.~\ref{eqn:eom.pressure.entropy.betterh}) formulations of the EOM, motivated by the approaches above that endeavor to enforce single-valued SPH pressures near contact discontinuities. We consider these methods in a wide range of idealized and more complex test problems, and show that they {\em simultaneously} maintain manifest conservation while tremendously improving the treatment of contact discontinuities and fluid mixing processes.

\vspace{-0.5cm}
\section{The SPH Lagrangian \&\ Equations of Motion}
\label{sec:eom}

\subsection{A Fully General Derivation}
\label{sec:derivation}

Following \Sentropy, note that the SPH equations can be derived self-consistently from the discrete particle Lagrangian 
\be
L({\bf q},\,{\bf \dot{q}}) = \frac{1}{2}\,\sum_{i=1}^{N}\,m_{i}\,{\bf \dot{r}}_{i}^{2} - \sum_{i=1}^{N}\,m_{i}\,u_{i}
\ee
in the independent variables ${\bf q} = ({\bf r}_{1},...,{\bf r}_{N},h_{1},...,h_{N})$, namely the positions and smoothing lengths of each volume element/particle, and the internal energy per unit mass $u$. If the smoothing lengths $h$ are constant, then the only independent variables are the ${\bf r}_{i}$ and the equations of motion follow from $d(\partial L/\partial \dot{q}_{i})/dt = \partial L/\partial q_{i}$. We require the derivatives of the $u_{i}$; recalling that this is at constant entropy, so ${\rm d}u=-(P/m)\,{\rm d}V$, we have:
\be
\label{eqn:dudv}
\frac{\partial u_{i}}{\partial q_{i}}{\Bigr|}_{A} = -\frac{P_{i}}{m_{i}}\,\frac{\partial \DV_{i}}{\partial q_{i}}
\ee
where $P_{i}$ is the pressure and $\DV_{i}$ is {\em some} estimator of the particle ``volume.'' 

We clearly require some thermodynamic variable to determine $P$; we can choose ``which'' to follow, for example internal energy $u$ or entropy $A$. For a gas which is polytropic under adiabatic evolution (we consider more general cases below), if we follow particle-carried $u_{i}$, then self-consistency with the thermodynamic equation above requires that we define the pressure in the above equation as $P_{i} = (\gamma-1)\,u_{i}\,(m_{i}/\DV_{i})$; if we follow the entropy $A_{i}$ we must define $P_{i} = A_{i}\,(m_{i}/\DV_{i})^{\gamma}$. 

If $h$ is allowed to vary, then we require some relation which makes it differentiable in order to make progress. This usually amounts to enforcing some condition on the effective ``neighbor number'' or ``mass inside a kernel.'' We stress that this language is somewhat misleading: it is {not} actually the case that there is exactly a certain neighbor number or mass inside the kernel, but rather that a continuous relation between $h$ and some local volumetric quantity is enforced, for example $(4\pi/3)\,h_{i}^{3}\,\rho_{i} = M_{\rm kernel} = m_{i}\,N_{\rm ngb}$ (so that $h_{i} \propto \rho_{i}^{-1/3}$). Any such constraint (if continuous) is equally valid: motivated by the ``effective neighbor number'' approach we can define the constraint equation
\be
\label{eqn:constraint}
\phi_{i}({\bf q}) \equiv \frac{4\pi}{3}\,h_{i}^{3}\,\frac{1}{\DVa_{i}} - N_{\rm ngb}= 0
\ee
where $\DVa_{i}$ is some continuous estimator of the ``particle volume''; e.g.\ for $\DVa_{i} = m_{i}/\rho_{i}$, we recover the approximate ``mass in kernel'' constraint. 

We stress that $\DVa$ does {\em not} need to be the same as $\DV$; one is the effective volume used to evolve the thermodynamics, the other is simply {\em any} continuous function used to define the $h_{i}$, so as to make them differentiable and thus allow us to include the appropriate ``$\nabla h$'' terms in the EOM. 

The equations of motion can then be determined from 
\be
\frac{{\rm d}}{{\rm d}t}\,\frac{\partial L}{\partial \dot{q}_{i}} - \frac{\partial L}{\partial q_{i}} = \sum_{j=1}^{N}\,\lambda_{j}\,\frac{\partial \phi_{j}}{\partial q_{i}}
\ee
where $\lambda_{i}$ are the Lagrange multipliers. 
The second half of these equations ($q_{i} = h_{i}$) lead to the Lagrange multipliers 
\begin{align}
\lambda_{i} &= -\frac{3\,P_{i}\,\DVa_{i}^{2}}{4\pi\,h_{i}^{3}}\,\psi_{i}\\
\psi_{i} &\equiv \frac{h_{i}}{3\,\DVa_{i}}\,\frac{\partial \DV_{i}}{\partial h_{i}}
{\Bigl [}1 - \frac{h_{i}}{3\,\DVa_{i}}\,\frac{\partial \DVa_{i}}{\partial h_{i}} {\Bigr]}^{-1}
\end{align}
Inserting this into the first half of the equations gives the EOM
\begin{align}
m_{i}\,\frac{{\rm d}{\bf v}_{i}}{{\rm d}t} = 
\sum_{j=1}^{N}\,P_{j}\,{\Bigl[}\nabla_{i}\DV_{j} + \psi_{j}\,\nabla_{i}\DVa_{j} {\Bigr]}
\end{align}

Now we require some way of defining ``volumes.'' In SPH this is done with respect to the kernel sum: for any particle-carried scalar value $x_{i}$, the $x$-weighted volume average $y_{i}=\bar{y}$ can be constructed as
\begin{align}
y_{i} &\equiv \sum_{j=1}^{N}\,x_{j}\,W_{ij}(h_{i}) 
\end{align}
where $W_{ij}(h_{i}) = W_{ij}(|r|_{ij}/h_{i})$ is the smoothing kernel (discussed further below) as a function of $|r|_{ij}\equiv |{\bf r}_{i}-{\bf r}_{j}|$. The corresponding $x$-weighted volume element can then be defined as $\DV_{i}\equiv x_{i}/y_{i}$. Note that for the choice $x_{i}=m_{i}$, we recover the familiar ``standard'' SPH choices of $y_{i} = \rho_{i}$ and $\DV_{i} = m_{i}/\rho_{i}$; but any other choice of $x_{i}$ is in principle equally valid. For any well-behaved kernel, this makes the $\DV_{i}$ fully differentiable functions of ${\bf r}$ and $h$: 
\begin{align}
\frac{\partial \DV_{i}}{\partial h_{i}} = -\frac{x_{i}}{y_{i}^{2}}\,\frac{\partial y_{i}}{\partial h_{i}},\ \ \ 
\nabla_{i}\DV_{j} = -\frac{x_{j}}{y_{j}^{2}}\,\nabla_{i}\, y_{j} 
\end{align}
with 
\begin{align}
\nabla_{i}\, {y}_{j} &= x_{i}\,\nabla_{i}\,W_{ij}(h_{j}) + \delta_{ij}\,\sum_{k=1}^{N}\,x_{k}\,\nabla_{i}\,W_{ik}(h_{i}) \\ 
\frac{\partial {y}_{i}}{\partial h_{i}} &= -\sum_{j=1}^{N}\,\frac{x_{j}}{h_{i}}\,{\Bigl(}3\,W_{ij}(h_{i}) + \frac{|r|_{ij}}{h_{i}}\,\frac{\partial W(|r|/h)}{\partial(|r|/h)}{\Bigr |}_{|r|_{ij}/h_{i}}{\Bigr)}
\end{align}

Putting all of this together, we obtain
\begin{align}
\label{eqn:eom}
m_{i}\,\frac{{\rm d}{\bf v}_{i}}{{\rm d}t} &= 
-\sum_{j=1}^{N}\,x_{i}\,x_{j}\,
{\Bigl[} \frac{P_{i}}{y_{i}^{2}}f_{ij}\,\nabla_{i}W_{ij}(h_{i})
+ \frac{P_{j}}{y_{j}^{2}}f_{ji}\,\nabla_{i}W_{ij}(h_{j}){\Bigr]} \\ 
f_{ij} &\equiv 1 - \frac{\tilde{x}_{j}}{x_{j}}\,{\Bigl(}\frac{h_{i}}{3\,\tilde{y}_{i}}\,\frac{\partial y_{i}}{\partial h_{i}}{\Bigr)}\,{\Bigl[}1 + \frac{h_{i}}{3\,\tilde{y}_{i}}\,\frac{\partial \tilde{y}_{i}}{\partial h_{i}} {\Bigr]}^{-1}
\end{align}


Because the ``potential'' (thermal energy) in the Lagrangian here depends only on coordinate differences and is rotationally symmetric, the pair-wise force in Eq.~\ref{eqn:eom} is automatically anti-symmetric (i.e.\ obeys Newton's third law); energy, entropy, momentum, and angular momentum are all {\em manifestly} conserved, provided that smoothing lengths are adjusted to ensure the appropriate $\phi$ constraint (it is straightforward to verify this explicitly). The so-called ``$\nabla h$'' terms, which depend on derivatives in time and space of the smoothing lengths, are implicitly included to all orders, via the $f$ terms. The final EOM for any choice of $x_{i}$ and $\tilde{x}_{i}$ involves essentially identical information, constraints, and cost. In other words, because all of the formulations we consider are simply replacing the choice of particle-carried scalar in Eq.~\ref{eqn:eom}, they involve identical computational expense for otherwise equal gas conditions.

\vspace{-0.5cm}
\subsection{Formulations of the SPH Equations}
\label{sec:formulations}

\subsubsection{Density-Entropy Formulation}
\label{sec:density.entropy}

If we take $x_{i}=\tilde{x}_{i}=m_{i}$ (giving $y_{i}=\tilde{y}_{i}= \bar{\rho}_{i}$, the kernel-averaged mass density estimator, and volume estimator $\DV_{i}=m_{i}/\rho_{i}$), and follow the entropy $A_{i}$ (giving $P_{i} = \bar{P}_{i}= A_{i}\,\bar{\rho}_{i}^{\gamma}$), we obtain the EOM from \Sentropy: 
\begin{align}
\label{eqn:eom.density.entropy}
\frac{{\rm d}{\bf v}_{i}}{{\rm d}t} 
&= -\sum_{j=1}^{N}\,m_{j}\,
{\Bigl[}
\frac{f_{i}\,\bar{P}_{i}}{\bar{\rho}_{i}^{2}}{\nabla_{i}W_{ij}(h_{i})} +
\frac{f_{j}\,\bar{P}_{j}}{\bar{\rho}_{j}^{2}}{\nabla_{i}W_{ij}(h_{j})} 
{\Bigr]} \\ 
\nonumber &= -\sum_{j=1}^{N}\,m_{j}\,
{\Bigl[}
{f_{i}\,A_{i}\,}{\bar{\rho}_{i}^{\gamma-2}}{\nabla_{i}W_{ij}(h_{i})} +
{f_{j}\,A_{j}\,}{\bar{\rho}_{j}^{\gamma-2}}{\nabla_{i}W_{ij}(h_{j})} 
{\Bigr]} \\
\nonumber f_{i} &= {\Bigl [}1 + \frac{h_{i}}{3\,\bar{\rho}_{i}}\,\frac{\partial \bar{\rho}_{i}}{\partial h_{i}}{\Bigr ]}^{-1},\ \ \ \ \bar{\rho}_{i} \equiv \sum_{j=1}^{N}\,m_{j}\,W_{ij}(h_{i})
\end{align}

Note that for adiabatic evolution, we require no energy equation, since entropy is followed; for a specific energy {\em defined} as $u \equiv \bar{P}_{i}/[(\gamma-1)\,\bar{\rho}_{i}] = (\gamma-1)^{-1}\,A_{i}\,\bar{\rho}_{i}^{\gamma-1}$, energy conservation is manifest from the EOM above.

As discussed in \S~\ref{sec:intro}, this formulation is known to have trouble treating certain contact discontinuities. Because the only volumetric quantity that enters is the density, this fails when the densities are no longer differentiable, even when pressure is smooth/constant. Specifically, consider a contact discontinuity, $\rho_{1}\,c_{1}^{2} = \rho_{2}\,c_{2}^{2}$ (where quantities `1' and `2' are on either side of the discontinuity). As we approach the discontinuity, the kernel-estimated density must trend to some average because it spherically averages over both ``sides,'' $\rho\rightarrow \langle \rho \rangle$, but the particle-carried sound speeds $c$ remain distinct, so the pressure is now multi-valued, with a `pressure blip' of magnitude $\sim (\rho_{\rm max}/\rho_{\rm min})\,P_{\rm true}$ appearing. This has a gradient across the central smoothing length, causing an artificial, repulsive ``surface tension'' force that suppresses interpenetration across the discontinuity.

\vspace{-0.5cm}
\subsubsection{Pressure-Energy Formulation}
\label{sec:pressure.energy}

Instead consider $x_{i}=\tilde{x}_{i} = (\gamma-1)\,U_{i}\equiv (\gamma-1)\,m_{i}\,u_{i}$, proportional to the particle internal energy. Now, $y_{i}=\tilde{y}_{i}$ is by definition a direct kernel-averaged pressure estimator $y_{i} = \bar{P}_{i}$, and $\DV_{i} = (\gamma-1)\,m_{i}\,u_{i}/P_{i}$. This means that the pressure itself is now the directly kernel-averaged quantity entering the EOM and is therefore always single-valued. So long as the pressure is smooth/differentiable (regardless of how the density varies), the EOM should be well-behaved. For this choice, we obtain:\footnote{Eq.~\ref{eqn:eom.pressure.energy} is similar to the EOM derived in \SM, itself identical to that derived earlier in \citet{ritchie.thomas:2001.egy.wtd.sph} from purely heuristic arguments. These do, after all, motivate the derivation here. However there are two key differences. First, we derive and include the $\nabla h$ terms ($f_{i}=1$ in \SM), necessary for conservation if $h$ varies. Second, the $\nabla W$ terms enter differently (with different multipliers and indices). This stems from the Lagrangian derivation and is necessary -- even for constant $h$ -- for the EOM to {\em simultaneously} conserve energy and entropy (i.e.\ to properly advect the thermodynamic volume element).}
\begin{align}
\nonumber \frac{{\rm d}{\bf v}_{i}}{{\rm d}t} &= -\sum_{j=1}^{N}\,(\gamma-1)^{2}m_{j}\,u_{i}\,u_{j}\,
{\Bigl[}
\frac{f_{i}}{\bar{P}_{i}}\,\nabla_{i}W_{ij}(h_{i}) + \frac{f_{j}}{{\bar{P}_{j}}}\,\nabla_{i}W_{ij}(h_{j})
{\Bigr]}  \\ 
 f_{i} &= {\Bigl [}1 + \frac{h_{i}}{3\,\bar{P}_{i}}\,\frac{\partial \bar{P}_{i}}{\partial h_{i}}{\Bigr ]}^{-1},\ \ \ \ \bar{P}_{i} \equiv \sum_{j=1}^{N}\,(\gamma-1)\,m_{j}\,u_{j}\,W_{ij}(h_{i})
\label{eqn:eom.pressure.energy}
\end{align}

Recall, we now need to evolve the energy explicitly. From Eq.~\ref{eqn:dudv}, ${\rm d}u/{\rm d}t = -(P/m)\,({\rm d}\DV/{\rm d}t)$, and the evolution of the volume element $\DV$ just follows from the Lagrangian continuity equation, so we obtain
\begin{align}
\frac{{\rm d}u_{i}}{{\rm d}t} = \sum_{j=1}^{N} (\gamma-1)^{2}\,m_{j}\,u_{i}\,u_{j}\,\frac{f_{i}}{\bar{P}_{i}}\,({\bf v}_{i}-{\bf v}_{j})\cdot\nabla_{i}W_{ij}(h_{i})
\end{align}
It is straightforward to see that this guarantees explicit energy conservation; entropy conservation is implicit and also easily verified.

There are, however, significant drawbacks to the choice of $\tilde{x}_{i}=x_{i}$ ($\DVa=\DV$), in which case we are implicitly defining $h$ such that $h_{i}^{3} \propto u_{i}/\bar{P}_{i}$. This is, by itself, perfectly valid and is easily solved by the same bisector method as in the ``standard'' ($\DVa=m/\rho$) formulation. However, in practice, the particle $u_{i}$ values vary much more widely than the $m_{i}$. This leads to some potential problems. First, if there is large variation in $u_{i}$, convergence in $h_{i}$ can become quite expensive. Second, again under circumstances with large disorder, the required $h_{i}$ can become very large, leading to an effective loss of resolution. Third and most problematic, under some circumstances the constraint $\phi$ can have multiple solutions; in this case if $h_{i}$ ``jumps,'' it is no longer continuously differentiable, and so exact energy conservation is broken.

An obvious alternative is to use $\tilde{x}\equiv1$, i.e.\ $\DVa_{i}\equiv 1/\bar{n}_{i}$, where $\bar{n}_{i}$ is the ``particle number density'' 
\be
\bar{n}_{i} = \tilde{y}_{i}(\tilde{x}=1) \equiv \sum_{j=1}^{N}W_{ij}(h_{i})
\ee
This restores the effective ``number of neighbors'' criterion for $h_{i}$, and is always well-behaved since all particles are weighted equally. If we do this, the EOM become:
\begin{align}
\nonumber \frac{{\rm d}{\bf v}_{i}}{{\rm d}t} &= -\sum_{j=1}^{N}\,(\gamma-1)^{2}m_{j}\,u_{i}\,u_{j}\,
{\Bigl[}
\frac{f_{ij}}{\bar{P}_{i}}\,\nabla_{i}W_{ij}(h_{i}) + \frac{f_{ji}}{{\bar{P}_{j}}}\,\nabla_{i}W_{ij}(h_{j})
{\Bigr]}  \\ 
f_{ij} &= 1 - {\Bigl(}\frac{h_{i}}{3(\gamma-1)\,\bar{n}_{i}\,m_{j}\,u_{j}}\,\frac{\partial \bar{P}_{i}}{\partial h_{i}}{\Bigr)}\,
{\Bigl[}{1 + \frac{h_{i}}{3\,\bar{n}_{i}}\,\frac{\partial \bar{n}_{i}}{\partial h_{i}}}{\Bigr]}^{-1}
\label{eqn:eom.pressure.energy.betterh}
\end{align}
Note that this is just the previous equation with $f_{i}\rightarrow f_{ij}$; in other words, the EOM are identical up to the ``$\nabla h$'' corrections, which is what we expect, since the only function of the $\DVa$ term is to determine how the $h_{i}$ evolve. Trivially, then, the energy equation is also the same as above but with $f_{i}\rightarrow f_{ij}$. 

As discussed in \SM, because the volumetric quantity used in the EOM here is now directly the kernel-estimated pressure (instead of the density), this formulation automatically {\em guarantees} that pressure is single-valued at contact discontinuities, and so removes the pressure ``blip'' and surface tension force. The equations will now be well-behaved so long as pressure is smooth. This is true by definition in contact discontinuities; it is of course not true at shocks, but neither (typically) is the density constant there -- so we do not lose any desirable behaviors of the density-entropy formulation. In either case, we require some artificial viscosity to treat shocks.

\vspace{-0.5cm}
\subsubsection{Pressure-Entropy Formulation}
\label{sec:pressure.entropy}

If we wish to retain a direct kernel-estimate of the pressure entering the EOM, but formulate this in terms of entropy, we must instead consider $x_{i}=m_{i}\,A_{i}^{1/\gamma}$. In this case,\footnote{This choice of $x_{i}$ may seem a bit strange, but in fact this is the only self-consistent ``entropy formulation'' which directly evaluates the pressure. If we simply substituted $u_{j}=A_{j}\,\bar{\rho}_{j}^{\gamma-1}/(\gamma-1)$ in $x_{i}= (\gamma-1)\,m_{i}\,u_{i}$, we would re-introduce the density $\bar{\rho}$ (which we are trying to avoid in this formulation of the equations); we could instead define $u_{j} = A_{j}\,(\bar{P}_{j}/A_{j})^{\gamma-1}/(\gamma-1)$, but this involves $\bar{P}_{j}$ in its own definition and would require a prohibitively expensive iterative solution over all particles every timestep.} we obtain by definition (from the consistency requirement for $P_{i}$) 
\be
\bar{P}_{i} = y_{i}^{\gamma} = {\Bigl [} \sum_{j=1}^{N}m_{j}\,A_{j}^{1/\gamma}\,W_{ij}(h_{i}) {\Bigr]}^{\gamma}
\ee

If we also assume $\tilde{x}_{i}=x_{i}$, the EOM become
\begin{align}
\nonumber \frac{{\rm d}{\bf v}_{i}}{{\rm d}t} &= -\sum_{j=1}^{N}\,m_{j}\,(A_{i}\,A_{j})^{\frac{1}{\gamma}}\,
{\Bigl[}
\frac{f_{i}\,\bar{P}_{i}}{\bar{P}_{i}^{2/\gamma}}\,\nabla_{i}W_{ij}(h_{i}) + \frac{f_{j}\,\bar{P}_{j}}{{\bar{P}_{j}^{2/\gamma}}}\,\nabla_{i}W_{ij}(h_{j})
{\Bigr]}  \\
f_{i} &= {\Bigl [}1 + \frac{h_{i}}{3\,\bar{P}_{i}^{1/\gamma}}\,\frac{\partial \bar{P}_{i}^{1/\gamma}}{\partial h_{i}}{\Bigr ]}^{-1}
\label{eqn:eom.pressure.entropy}
\end{align}

Note however that using $\tilde{x}_{i}=x_{i}$ implies $\DVa_{i}=m_{i}\,(A_{i}/\bar{P}_{i})^{1/\gamma}$; as in the previous section, this can introduce problems of convergence and diffusion. Therefore, instead consider as before $\tilde{x}_{i}=1$ ($\DVa_{i} = 1/\bar{n}_{i}$). This gives:
\begin{align}
\frac{{\rm d}{\bf v}_{i}}{{\rm d}t} &= -\sum_{j=1}^{N}\,m_{j}\,(A_{i}\,A_{j})^{\frac{1}{\gamma}}\,
{\Bigl[}
\frac{f_{ij}\,\bar{P}_{i}}{\bar{P}_{i}^{2/\gamma}}\,\nabla_{i}W_{ij}(h_{i}) + \frac{f_{ji}\,\bar{P}_{j}}{{\bar{P}_{j}^{2/\gamma}}}\,\nabla_{i}W_{ij}(h_{j})
{\Bigr]}  
\label{eqn:eom.pressure.entropy.betterh}
\end{align}
with
\begin{align}
f_{ij} &= 1-{\Bigl(}\frac{h_{i}}{3\,A_{j}^{1/\gamma}\,m_{j}\,\bar{n}_{i}}\,\frac{\partial \bar{P}_{i}^{1/\gamma}}{\partial h_{i}}{\Bigr)}\,{\Bigl[}{1 + \frac{h_{i}}{3\,\bar{n}_{i}}\,\frac{\partial \bar{n}_{i}}{\partial h_{i}}}{\Bigr]}^{-1}
\end{align}
As in the density-entropy formulation, we explicitly evolve the entropy so for adiabatic evolution require no additional evolution equation.

This formulation is very similar to the pressure-energy formulation, (and has the identical advantages of good behavior at contact discontinuities). The only difference is the free choice of thermodynamic variable. This formulation trivially conserves entropy, and manifestly conserves energy to machine differencing accuracy if constant timesteps are used (the choice of pressure-energy or pressure-entropy formulation can lead to some differences when adaptive timesteps are used, but we show these are generally small). It is largely a matter of convenience and minor computational expense  which method is preferred. 

When the pressure is smooth and there is good particle order, the $f_{ij}\approx1$ here, which means our choice of how to regularize $h$ is unimportant, and no spurious ``surface tension'' force is introduced. For the choice $\tilde{x}=1$, the correction terms remain well-behaved even if there is large particle disorder in $A_{i}$, critical to stability in simulations when heating/cooling are included and entropy is no longer conserved. Another useful feature here is the following: imagine the case where there is large particle disorder so $P_{j}\gg P_{i}$ and $A_{j}\gg A_{i}$. Since the $A$ terms enter as multiplicative pre-factors, their difference does not introduce errors into the sum; gradient errors will arise from differencing the $P_{i}$ terms, but for $\gamma=5/3$, these enter only as $P_{i}^{-1/5}$, so differencing errors are greatly suppressed.

\vspace{-0.4cm}
\subsubsection{More General Cases}
\label{sec:general.eos}

In \S~\ref{sec:density.entropy}-\ref{sec:pressure.entropy}, we simplify by assuming the gas obeys a polytropic equation of state under differential adiabatic compression or expansion. We emphasize that this does not exclude the gas undergoing shocks (in which the entropy and energy change according to artificial viscosity), cooling, and/or chemical evolution (additional operations to ${\rm d}u_{i}$ in Eq.~\ref{eqn:dudv}); these are just handled in an additional, separate step or loop each timestep (see an example in \S~\ref{sec:feedback.example}). 

However, some situations call for more complicated equations of state. Consider the case where the pressure of a given particle $P_{i}$ is an arbitrarily complicated (but single-valued) function $g$ of the thermodynamic volume element $\DV_{i}$ and the local (particle-carried) state variables ${\bf {a}}_{i} = (a_{i,\,1},\,a_{i,\,2},...,a_{i,\,m})$, so $P_{i} = g(\DV_{i},\,{\bf {a}}_{i})$. The ${\bf {a}}$ might include $m_{i}$ and $u_{i}$ or $A_{i}$, as in our previous examples, but also information about the chemical state, radiation field, position or velocity, phase, etc, of the gas. 
Our general form of the EOS in Eq.~\ref{eqn:eom} made no assumption about the equation of state, and still holds. The question is how to determine the appropriate $x_{i}$ and $\tilde{x}_{i}$ for a ``pressure formulation.'' This requires any $x_{i}$ such that there is a one-to-one mapping between the smoothing kernel sum and the pressure (so that $\nabla (\DV)$ vanishes when $\nabla P$ does). We can ensure this by choosing $x_{i}$ to be the solution to the equation $g(\DV_{i} = x_{i},\,{\bf a}_{i}) = 1$ (i.e.\ if we were to replace $\DV_{i}$ by $x_{i}$, which we recall typically has different units, we would obtain a dimensionless $P_{i}=1$). We then define $\DV_{i} = x_{i}/y_{i} = x_{i}/\sum x_{j}\,W_{ij}(h_{i})$ as usual, and take $P_{i} = g(\DV_{i} = x_{i}/y_{i},\,{\bf a}_{i})$ in Eq.~\ref{eqn:eom}. We still have full freedom to define $\tilde{x}_{i}$; based on the formulations above, we suggest that the simple choice $\tilde{x}_{i}=1$ is often most stable.

It is straightforward to verify that our choices in both the pressure-energy ($P_{i}=(\gamma-1)\,u_{i}\,m_{i}/\DV_{i}$) and pressure-entropy ($P_{i}=A_{i}\,(m_{i}/\DV_{i})^{\gamma}$) formulations satisfy the condition $g(\DV_{i}=x_{i},\,{\bf a}_{i})=1$. These are just special cases of the above with ${\bf a}_{i}=(m_{i},\,u_{i})$ or ${\bf a}_{i}=(m_{i},\,A_{i}$) (and different $g$), respectively.

A more complicated example might be a case with a chemical potential, such that $m_{i}\,{\rm d}u_{i} |_{A} = -P_{i}\,{\rm d}\DV_{i} + \sum \mu_{k}\,{\rm d} N_{k}$ where the $k$ are different species. If the $N_{k}$ are constant over a differential adiabatic volume change, then this can be treated by any of the above formulations with the chemistry evolved (under the effects of radiation, for example) in a separate step. If the $N_{k}$ are functions of the volume element $\DV$, however, then we simply have $m_{i}\,{\rm d}u_{i} |_{A} \rightarrow -\hat{P}_{i}\,{\rm d}\DV_{i}$ where $\hat{P}_{i} \equiv P_{i} - \sum \mu_{k}\,(\partial N_{k} / \partial \DV_{i})$. So we can use the EOM in Eq.~\ref{eqn:eom} with $P_{i}\rightarrow \hat{P}_{i}$, and use the approach above to determine the appropriate $x_{i}$ (with $\tilde{x}_{i}=1$).

\vspace{-0.5cm}
\section{Additional Simulation Ingredients}
\label{sec:sims}

\subsection{Entropy \&\ Artificial Viscosity}
\label{sec:sims:viscosity}

As is standard in SPH, the algorithm is inherently inviscid and some artificial viscosity must be included to properly capture shocks. However, we include a more sophisticated treatment of the artificial viscosity term (as compared to \Sentropy\ and ``standard'' {\small GADGET}; which follow \citealt{gingold.monaghan:1983.artificial.viscosity}), following \citet{morris:1997.sph.viscosity.switch} with a \citet{balsara:1989.art.visc.switch} switch. This includes a particle-by-particle artificial viscosity that grows rapidly in strong shocks and rapidly decays away from shocks (to a minimum $\alpha=0.05$), reducing numerical dissipation by more than an order of magnitude away from shocks compared to the previous constant artificial viscosity prescription. For detailed comparison of the viscosity algorithms, we refer to \citet{cullen:2010.inviscid.sph}.

\vspace{-0.5cm}
\subsection{Thermodynamic Evolution \&\ Timestep Criteria}
\label{sec:sims:timestep}

For all problems discussed here, we employ the {\small GADGET} adaptive timestep algorithm, which dramatically reduces the computational expense for almost all interesting problems (relative to using a constant simulation timestep). However, as pointed out in \citet{saitoh.makino:2009.timestep.limiter} and developed further in \citet{durier:2012.timestep.limiter}, in problems with very high Mach number shocks/bulk flows, the standard adaptive timestepping can lead to problems if particles with long timesteps interact suddenly mid-timestep with material evolving on much shorter timesteps. Fortunately this is easily remedied, and we do so by implementing a timestep limiter identical to that in \citet{durier:2012.timestep.limiter}. At all times, any active particle informs its neighbors of its timesteps and none are allowed to have a timestep $>4$ times that of a neighbor; and whenever a timestep is shortened (or energy is injected in feedback) particles are ``activated'' and forced to return to the timestep calculation as soon as possible.

\vspace{-0.5cm}
\subsection{Smoothing Kernel}

Our derivation of the EOM allows for an arbitrary choice of SPH smoothing kernel $W$, so long as it is differentiable. This choice can have a significant effect on some test problems via its effect on the resolution pressure gradient errors \citep[see e.g.][and discussion below]{morris:1996.sph.stability,dilts:1999.sph.stability,read:2010.sph.mixing.optimization}. We have experimented with a wide range of possible kernel shapes, following the $\sim10$ discussed in \citet{fulk.quinn:1996.sph.kernels} and \citet{hongbin.xin:05.sph.kernels}, the triangular kernels proposed in \citet{read:2010.sph.mixing.optimization}, and the variant Wendland kernels proposed in \citet{dehnen.aly:2012.sph.kernels}. However our intention here is not to study SPH kernels; for simplicity we therefore adopt a standard quintic spline kernel with $\NNb=128$ neighbors in all tests shown here (unless otherwise noted). This is the `optimal' spline kernel suggested in both \citet{hongbin.xin:05.sph.kernels} and \citet{dehnen.aly:2012.sph.kernels} and has effective resolution equal to a cubic spline with $34$ neighbors (but significantly higher accuracy). In most cases, we obtain similar results using a lower-order cubic spline\footnote{It is well-known that the ``standard'' \citet{schoenberg:1946.smoothing.kernels} spline kernels become unstable for arbitrarily high neighbor number; the ``correct'' way to increase $\NNb$ to better sample the kernel and reduce gradient errors is to move to higher-order kernels of $\mathcal{O}(n)\approx-1+(1.0-1.2)\,\NNb^{1/3}$. Done properly, this allows increasing $\NNb$ and accuracy without decreasing the effective spatial resolution (albeit at additional computational expense).} with $\NNb=32$; but we discuss where this is not the case. We do not see qualitative improvement in the specific tests here using yet higher-order kernels and/or increasing neighbor number as high as $\NNb\approx500$.

\vspace{-0.5cm}
\subsection{Density Estimation in Density-Independent SPH}
\label{sec:density.estimator}

An estimate of the density is often required for other calculations (e.g.\ cooling), even if it is not needed for the EOM. This can still be directly estimated from the standard SPH kernel, as $\rho_{i}$ above (i.e.\ a kernel sum with $x_{i}=m_{i}$). But in principle a density could also be inferred or estimated as $\rho_{th} \approx m_{i}/\DV_{i}$ for the thermodynamic $\DV$ ``pressure formulations'' (i.e.\ from the combination of the variables $\bar{P}_{i}$ and $A_{i}$ or $u_{i}$). However, in ``mixed'' regions near contact discontinuities, this will lead to multi-valued densities in neighboring particles (since entropies are still conserved at the particle level). This relates to the fact that the two densities represent physically distinct quantities. The former, direct kernel $\rho_{i}$ ($x_{i}=m_{i}$) is simply the volume-average mean mass density at the particle location. The latter (thermodynamically-inferred $\rho_{th}$) is the energy or entropy-weighted mean mass density (in the pressure-energy or pressure-entropy formulations, respectively), averaged over all neighbors within the kernel. Because the cooling is typically calculated on a particle-by-particle basis, we find the former estimator is more stable and appropriate in most applications. However, there may be situations where alternative weightings for the density estimator are optimal, and a more complete treatment of mixing may require a mechanism for equilibrating entropies inside the kernel and generating mixing entropy.

\begin{figure*}
    \plotsidesize{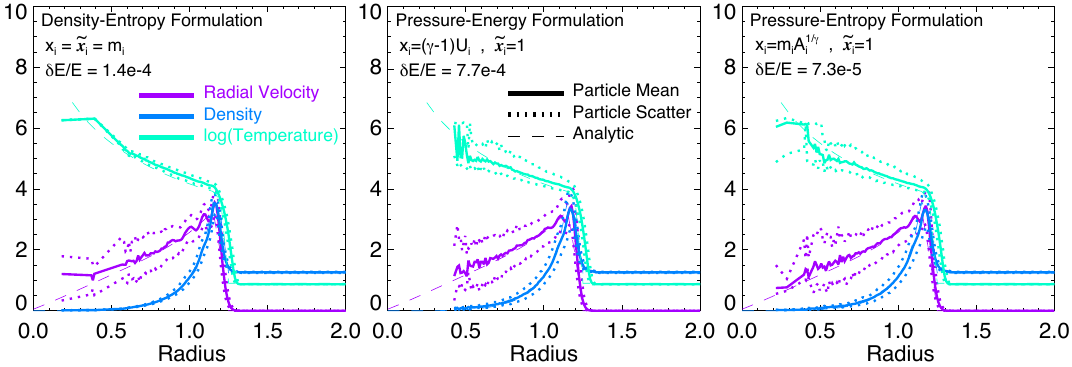}{0.98}
    \caption{Strong Sedov-Taylor blastwave test. We compare the solution at fixed time and resolution, with different SPH algorithms discussed in the text. 
    We plot radial velocity, density, and temperature (for clarity, we plot the median and $1\%-99\%$ interval of particle values binned in radial intervals $\Delta r=0.01$), with the analytic solution. We measure the accuracy of energy conservation $\delta E/E \equiv |E(t) - E(t=0)|/E(t=0)$. 
    {\em Left:} Standard SPH (density-entropy; $x_{i}=\tilde{x}_{i}=m_{i}$); Eq.~\ref{eqn:eom.density.entropy}. 
    {\em Center:} Lagrangian Pressure-Energy formulation ($x_{i}=(\gamma-1)\,U_{i}$) with particle-number density based smoothing lengths ($\tilde{x}_{i}=1$); Eq.~\ref{eqn:eom.pressure.energy.betterh}. 
    {\em Right:} Lagrangian Pressure-Entropy formulation ($x_{i}=m_{i}\,A_{i}^{1/\gamma}$) with particle-number density based smoothing lengths ($\tilde{x}_{i}=1$); Eq.~\ref{eqn:eom.pressure.entropy.betterh}.
    Conservation and accuracy is excellent up to the resolution limits (lack of particles dominates the noise at small radii), in all three algorithms. The pressure formulations slightly increase the particle scatter in temperature, but reduce the post-shock ``ringing'' of the velocity solution. Conservation errors in all three cases are overwhelmingly dominated by the adaptive timesteps, not the EOM.
    \label{fig:sedov1}}
\end{figure*}

\begin{figure*}
    \plotsidesize{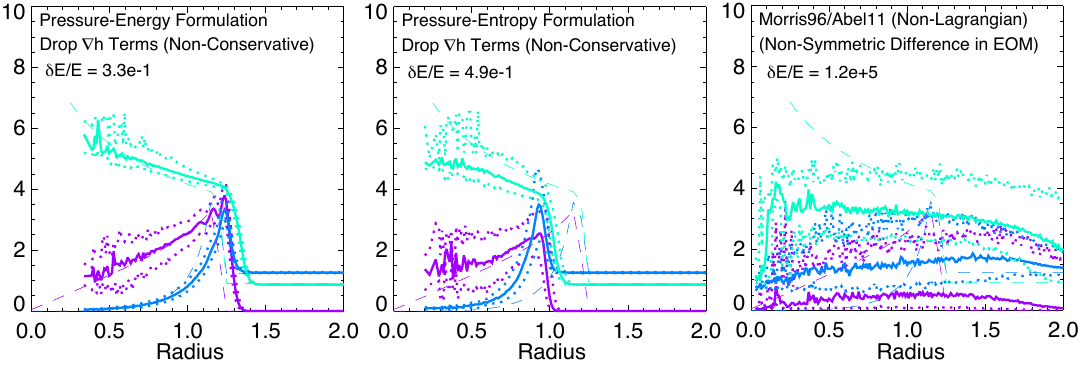}{0.95}
    \caption{As Fig.~\ref{fig:sedov1}, but for SPH equations that are not explicitly conservative. 
    {\em Top:} Pressure-Energy formulation {\em without} the $\nabla h$ terms ($f_{i}=1$ in Eq.~\ref{eqn:eom.pressure.energy}). Variations in $h$ now cause order-unity energy conservation errors, leading to the shock being in the wrong position. 
    {\em Middle}: Pressure-Entropy formulation without the $\nabla h$ terms. Again order-unity conservation errors appear and the shock evolves incorrectly. 
    {\em Bottom:} Non-Lagrangian SPH as considered in \citet{morris:1996.sph.stability,abel:2011.sph.pressure.gradient.est}; this algorithm minimizes linear pressure errors and removes the surface tension term, but violates conservation of energy and momentum. Conservation errors grow exponentially and dominate the solution. 
    \label{fig:sedov2}}
\end{figure*}

\begin{figure}
    \plotonesize{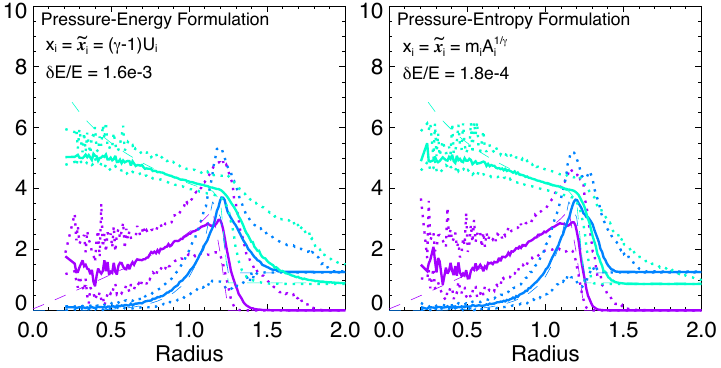}{1.02}
    \caption{As Fig.~\ref{fig:sedov1}, with explicitly conservative equations, but different choices of $\tilde{x}_{i}$ (how to regularize the $h_{i}$). 
    {\em Top:} Lagrangian Pressure-Energy formulation with $\tilde{x}_{i}=x_{i}$ (Eq.~\ref{eqn:eom.pressure.energy}). Although conservation is maintained, this choice of $\tilde{x}_{i}$ leads to enormous numerical diffusion and particle disorder (``spreading'' the shock location over a large radius and affecting the pre-shock gas). 
    {\em Bottom:} Lagrangian Pressure-Entropy formulation with $\tilde{x}_{i}=x_{i}$ (Eq.~\ref{eqn:eom.pressure.entropy}). Again, this leads to severe diffusion.
    \label{fig:sedov3}}
\end{figure}

\vspace{-0.5cm}
\section{Test Problems}

\subsection{Strong Sedov-Taylor Blastwaves}

Here we consider an extreme Sedov-Taylor blastwave, with very large mach number, designed to be a powerful test of conservation. A box of side-length $6\,$kpc is initially filled with $128^{3}$ equal-mass particles at constant density $n=0.5\,{\rm cm^{-3}}$ and temperature $10\,$K; $6.78\times10^{46}\,$J of energy is added to the central $64$ particles in a top-hat distribution. This triggers a blastwave with initial Mach number $\sim1000$, which we compare at $20\,$Myr, where the shock front should be at $r\approx1.19\,$kpc. In this test and below, unless otherwise specified, we assume a $\gamma=5/3$ gas equation of state.

First we compare the ``fully conservative'' algorithms we derive above. In ``standard'' SPH (density-entropy formulation; Eq.~\ref{eqn:eom.density.entropy}), the analytic solution is reproduced very well up to the SPH smoothing/resolution limit. The narrow shock jump is not perfectly resolved, for example, but averaged over the same smoothing, the agreement is excellent between analytic solution and mean particle values. There is also some scatter in particle properties (most notably velocity), but it is generally narrow. The largest deviation from the analytic solution is in the post-shock ``ringing'' in the velocity field, a well-known effect (which is sensitive to the artificial viscosity prescription). Note that the behavior at small radii $\lesssim0.5\,$kpc is noisy simply because the extremely low density means there are no particles here. As expected, the conservation properties are excellent as well; energy is conserved to a part in $\sim10^{4}$; similar conservation obtains for entropy, energy, momentum, and angular momentum. In fact the conservation errors are overwhelmingly dominated by the adaptive time-stepping scheme (which violates conservation by not always evolving mutually interacting particles at the same timesteps), not the formulation of the EOM. If we simply force constant, global timesteps, the conservation errors are reduced to machine accuracy (however this comes at great numerical expense).\footnote{We do confirm the point discussed in \S~\ref{sec:sims:timestep}, that motivates our timestepping algorithm: without care in ``signaling'' when adaptive timesteps are taken, conservation errors can be severe.}

To compare, the pressure-energy formulation ($x_{i}=(\gamma-1)\,U_{i}$ as Eq.~\ref{eqn:eom.pressure.energy.betterh}, using the neighbor number density volume element $\tilde{x}_{i}=1$ to define $h$) agrees very well, and also gives excellent conservation. The particle noise/scatter is slightly larger, most noticeably in the post-shock temperature. This occurs because of some ``mixing'' of thermal properties in the kernel average from higher-energy particles. However the post-shock ringing in the velocity solution (although still present) is reduced. 

The pressure-entropy formulation ($x_{i}=m_{i}\,A_{i}^{1/\gamma}$ with $\tilde{x}_{i}=1$, Eq.~\ref{eqn:eom.pressure.entropy.betterh}) is essentially identical to the pressure-energy formulation here (the slightly better conservation owes to how the sound speeds enter the timestep signaling scheme). This is not surprising -- the two represent essentially identical information but just choose to explicitly follow different thermodynamic variables. 

As discussed in \S~\ref{sec:derivation}, the EOM for the pressure-energy and pressure-entropy formulations take a simpler form if we assume $\tilde{x}_{i}=x_{i}$, as in standard SPH, but this can lead to problems. We show that here, by considering these implementations (Eq.~\ref{eqn:eom.pressure.energy} \&\ \ref{eqn:eom.pressure.entropy} for the pressure-energy and pressure-entropy formulations, respectively). Recall, these are still fully conservative Lagrangian formulations; as a result we still see good energy conservation. However, forcing the smoothing lengths to evolve not just with particle number density but with local thermodynamic quantities leads to occasional enormous ``super-smoothing'' that is both computationally expensive and introduces enormous numerical diffusion. We see that here, where the upper envelope of particles and even the mean are biased in the pre-shock medium (information from the post-shock region has clearly been over-smoothed into the regions here). Although some mean and post-shock quantities are well-behaved, this algorithm has introduced far too much numerical mixing. 

It is also instructive to see what happens if we drop the $\nabla h$ terms in these formations (returning to $\tilde{x}_{i}=1$). The EOM will now violate manifest energy conservation to the level that the smoothing lengths vary over an individual smoothing length (in a smooth medium, this correction should vanish at infinite resolution, but it will never vanish in shocks if variable $h$ are allowed). Indeed we now see order-unity energy errors. Although the qualitative solution appears similar, the pressure-energy and pressure-entropy solutions gain and lose energy, respectively, and so the shock is simply in the wrong position. 
This occurs in other, similar non-conservative formulations as well, for example see the discussion of problems with \citet{ritchie.thomas:2001.egy.wtd.sph} \&\ `OSPH' methods in \citet[][Appendix A]{read:2012.sph.w.dissipation.switches}.

Finally, we consider an algorithm which is manifestly non-conservative. Specifically, we consider the \citet{morris:1996.sph.stability} formulation of the pressure derivative, where the kernel sum for the EOM is over $(P_{j}-P_{i})/(\rho_{i}\,\rho_{j})\,\nabla W_{ij}$. This is very similar (although not identical) to the EOM proposed in \citet{abel:2011.sph.pressure.gradient.est} as well. It is possible to show that this formulation actually eliminates the leading-order gradient errors associated with the ``standard'' SPH EOM \citep[see e.g.][]{price:2012.sph.review}. However, the cost is manifest resolution-level violation of conservation (of energy, momentum, and entropy). If we adopt our standard initial conditions with this algorithm, we see that the conservation errors grow exponentially and quickly swamp the real solution. The problem, as described in \citet{price:2012.sph.review} is in the non-linear terms that maintain particle order in SPH.\footnote{\citet{abel:2011.sph.pressure.gradient.est} do show that the behavior of Sedov blastwaves in their formulation is tremendously improved if the initial conditions are modified so that the blastwave does not start from a point injection or top-hat particle distribution, but from an {\em already-developed} smaller blastwave with pre-initialized, resolved pressure gradients. However that is not the particular test here, and -- as they caution -- is not often the case in astrophysical systems where e.g.\ early-stage SNe explosions are unresolved.}

\begin{figure}
    \plotonesize{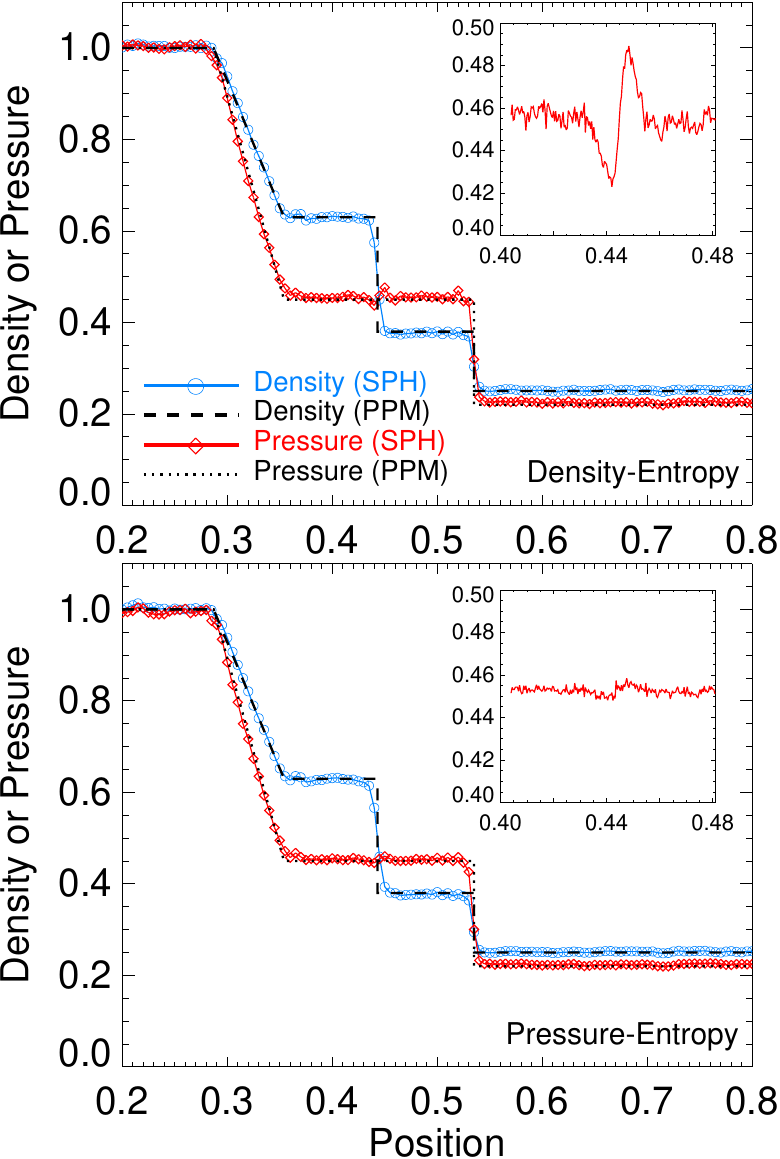}{0.95}
    \caption{Sod shock tube in three dimensions at time $t=0.1$. We show the median particle density and pressure along the long $x$-axis position (binned as Fig.~\ref{fig:sedov1}), compared to the reference solution from a piecewise parabolic method (PPM) solver. Inset shows the pressure profile binned in much smaller intervals around the location of the contact discontinuity ($x\approx0.44$), where the pressure should be constant. {\em Top:} Standard (density-entropy) SPH. The exact solution is generally reproduced well up to SPH smoothing, as Fig.~\ref{fig:sedov1}, but a ``pressure blip'' with an artificial gradient appears near the contact discontinuity. {\em Bottom:} Pressure-entropy SPH (Eq.~\ref{eqn:eom.pressure.entropy.betterh}, with $\tilde{x}_{i}=1$ for the reasons in Fig.~\ref{fig:sedov3}). The pressure blip is reduced to particle noise-level, while the agreement elsewhere with the PPM result remains.
    \label{fig:shocktube}}
\end{figure}

\vspace{-0.5cm}
\subsection{Sod Shock Tube}

Fig.~\ref{fig:shocktube} shows the results of a standard three-dimensional Sod shock tube test. We initialize a period domain with lengths $2,\,1/8,\,1/8$ in the $x,\,y,\,z$ directions, $5\times10^{4}$ particles, $\gamma=5/3$, zero initial velocities, densities $\rho=1,\,0.25$ and pressures $P=1,\,0.22$ in the left/right halves of the $x$ domain. In Fig.~\ref{fig:shocktube} we compare the results from both standard density-entropy SPH and the pressure-entropy formulation (with $\tilde{x}_{i}=1$) at time $t=0.1$. The median particle values agree well with the exact solution in both cases. As expected and seen in the Sedov case, the density and pressure discontinuities are smoothed by the SPH smoothing and artificial viscosity but otherwise well-behaved. The pressure discontinuity is slightly more smoothed by direct kernel averaging in the pressure-entropy case, but the difference is small. 

If we zoom in on the pressure profile near the contact discontinuity at $x\approx0.44$ (where the exact solution is $P=$\,constant), we see the ``pressure blip'' discussed in \S~\ref{sec:density.entropy} appear in the density-entropy case. The presence of the smoothed density in the EOM leads to an artificial pressure gradient; this occurs over a couple of SPH smoothing lengths with a fractional amplitude of $\sim10\%$. However, it is clearly above the particle-noise threshold, and we show below that it has significant effects. In the pressure-entropy formulation, this is almost completely eliminated (at least reduced to the noise level), whether we choose $\tilde{x}_{i}=1$ or $\tilde{x}_{i}=x_{i}$.

We have also repeated the 1D shock tube tests in \SM; our results are generally indistinguishable from their Figs.~2-4. 

\begin{figure}
    \plotonesize{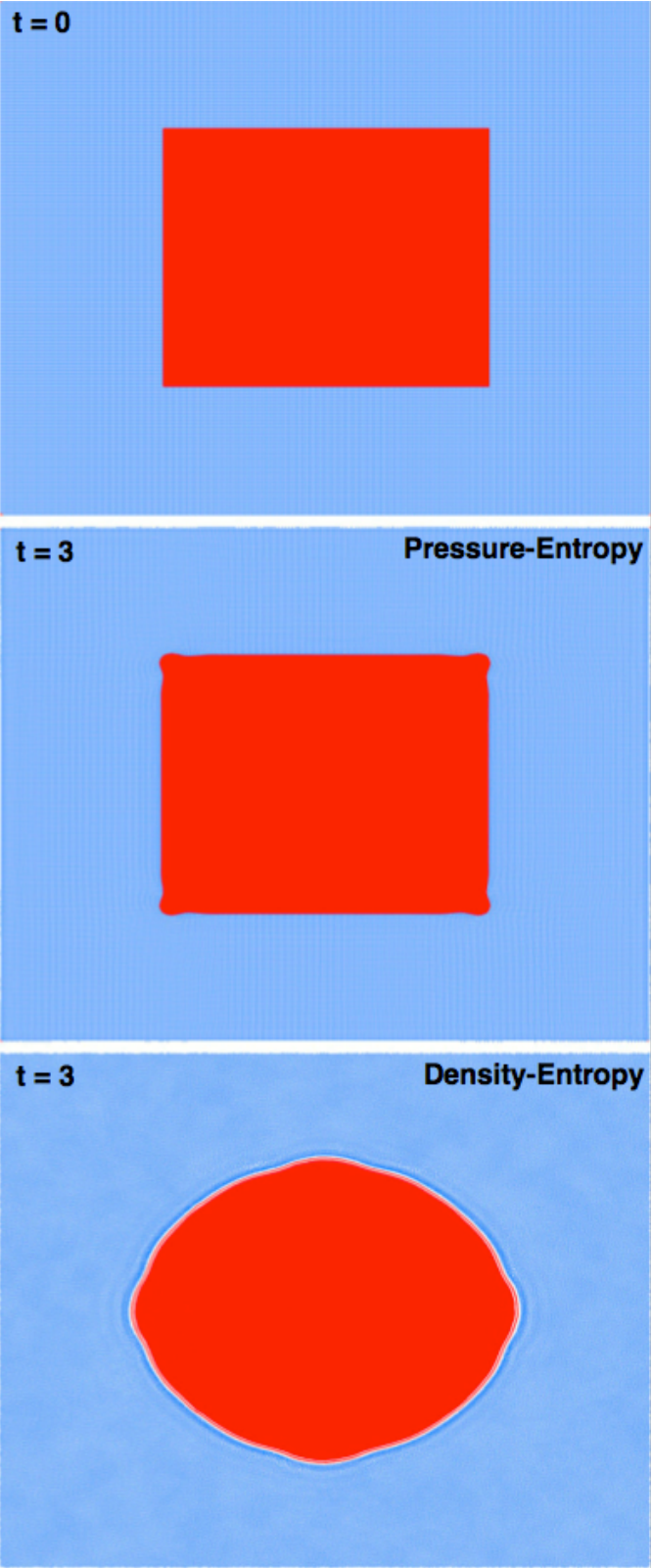}{0.8}
    \caption{Density (red $=4x$ blue) in a hydrostatic equilibrium test, with uniform pressure and no external forces.
    {\em Top:} Initial condition.
    {\em Middle:} System evolved to time $t=3$ in the pressure-entropy formulation (Eq.~\ref{eqn:eom.pressure.entropy.betterh}). The square evolves stably (the small corner rounding stems from the smoothing kernel).
    {\em Bottom:} Time $t=3$ in standard (density-entropy) SPH. The ``pressure blip'' around the contact discontinuity (Fig.~\ref{fig:shocktube}) manifests as an effective surface tension, opening a smoothing-length gap between the two fluids, and gradually deforming the square into a circle. 
    \label{fig:square}}
\end{figure}

\vspace{-0.5cm}
\subsection{Hydrostatic Equilibrium/Surface Tension Test}

Here we consider a simple test following \citet{cha:2010.godunov.sph,hes:2010.tesselation.sph} and \SM, that allows us to see the consequences of the ``surface tension'' effect discussed in \S~\ref{sec:intro}. We initialize a two-dimensional fluid in a square of length $L=1$ (with period boundaries) and constant pressure $P=3.75$, polytropic $\gamma=5/3$, and density $\rho=4\,\rho_{0}$ within a central square of length $L=1/2$ and $\rho=\rho_{0}=7/4$ outside. We use $256^{2}$ total particles, though the results are similar for as few as $\sim50^{2}$. 

Fig.~\ref{fig:square} shows the resulting system at $t=0$, and evolved to $t=3$, in the ``standard'' density-entropy formalism (Eq.~\ref{eqn:eom.density.entropy}) and pressure entropy-formulation (Eq.~\ref{eqn:eom.pressure.entropy.betterh}; it makes little difference for this test whether we adopt $\tilde{x}_{i}=x_{i}$ or $\tilde{x}_{i}=1$). This should be a stable configuration. But the density-entropy case behaves as a system with physical surface tension; a repulsive force appears on either side of the contact discontinuity from the ``pressure blip,'' opening the gap in the plot which then deforms the square to minimize the surface area of the contact discontinuity. It converges to a stable circle after a few relaxation oscillations. In the pressure-entropy case, the square remains stable for the duration of the runs we consider ($t=50$). There is a slight ``rounding'' of the corners, but this occurs quickly ($t<1$) then stabilizes; it appears to be a direct consequence of the smoothing kernel. These results are identical to those in \SM, for the pressure-energy formulation (with no $\nabla h$ terms).

\vspace{-0.5cm}
\subsection{Kelvin-Helmholtz Instabilities}

We next consider a (three-dimensional) Kelvin-Helmholtz (KH) test. The initial conditions are taken from the Wengen multiphase test suite\footnote{Available at \wengenurl} and described in \citet{agertz:2007.sph.grid.mixing,read:2010.sph.mixing.optimization}. Briefly, in a periodic box with size $256,\,256,\,16\,{\rm kpc}$ in the $x,\,y,\,z$ directions (centered on $0,\,0,\,0$), respectively, $\approx10^{6}$ equal-mass particles are initialized in a cubic lattice, with density, temperature, and $x$-velocity $=\rho_{1},\,T_{1},\,v_{1}$ for $|y|<4$ and $=\rho_{2}\,T_{2},\,v_{2}$ for $|y|>4$, with $\rho_{2}=0.5\,\rho_{1}$, $T_{2}=2.0\,T_{1}$, $v_{2}=-v_{1}=40\,{\rm km\,s^{-1}}$. The values for $T_{1}$ are chosen so the sound speed $c_{s,\,2}\approx 8\,|v_{2}|$; the system has constant initial pressure. To trigger instabilities, a sinusoidal velocity perturbation is applied to $v_{y}$ near the boundary, with amplitude $\delta v_{y} = 4\,{\rm km\,s^{-1}}$ and wavelength $\lambda=128\,{\rm kpc}$. 

\begin{figure}
    \plotone{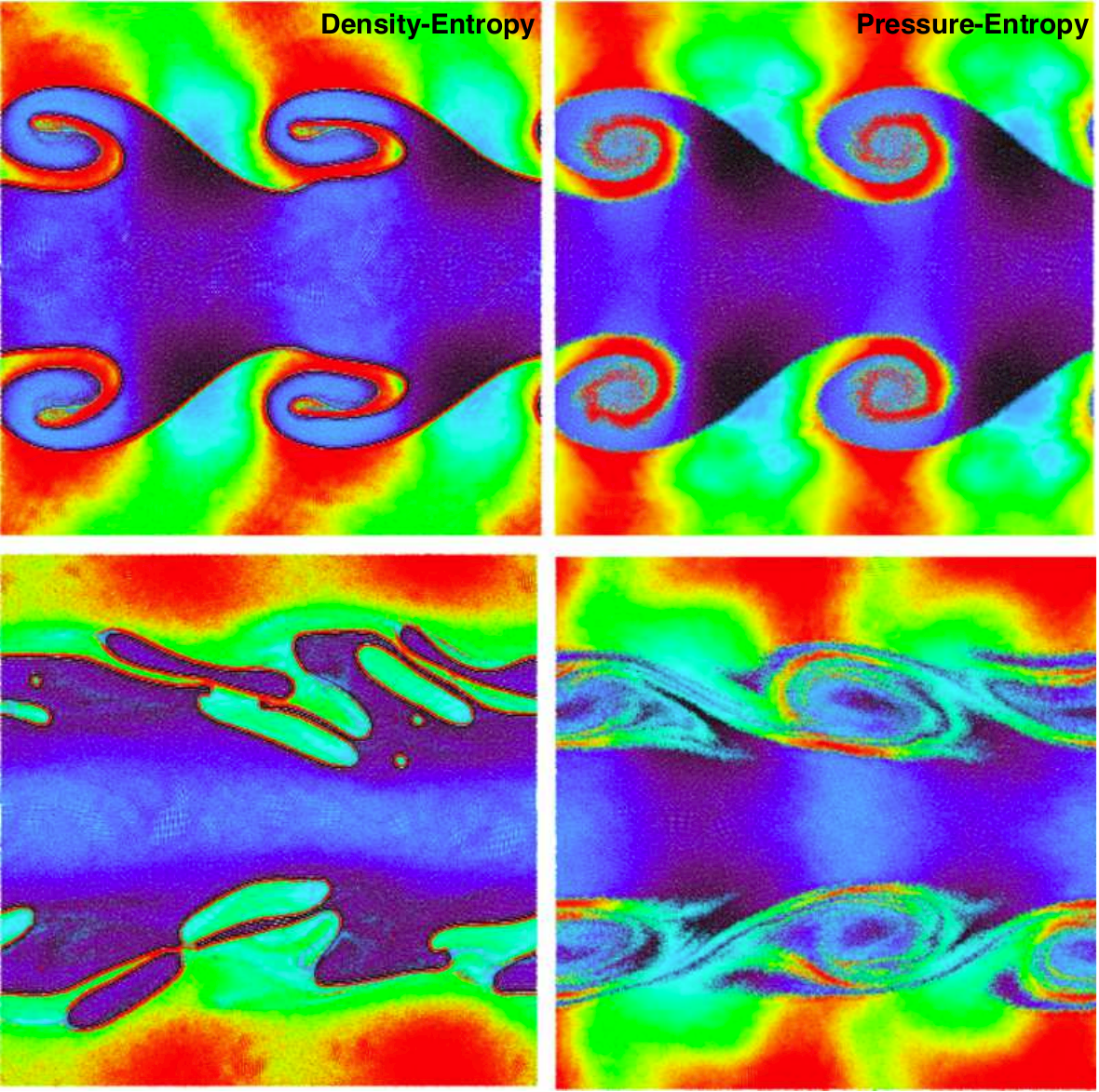}
    \caption{Specific entropy map of Kelvin-Helmholtz instabilities at time $2$ times the linear K-H growth timescale $\tau_{\rm KH}$ ({\em top}), 
    and at $8\,\tau_{\rm KH}$ ({\em bottom}). 
    {\em Left:} Standard (density-entropy) SPH. Note the ``surface layer'' surrounding the curls and the breakup into oil-and-water style blobs at later times. 
    {\em Right:} Pressure-entropy formulation ($\tilde{x}_{i}=1$). 
    \label{fig:kh1}}
\end{figure}

Fig.~\ref{fig:kh1} compares the resulting behavior in the ``standard'' density-entropy formulation of SPH (Eq.~\ref{eqn:eom.density.entropy}), and the pressure-entropy formulation (Eq.~\ref{eqn:eom.pressure.entropy.betterh}) proposed here. Because of the severe diffusion seen in the Sedov test associated with the choice of $x_{i}=\tilde{x}_{i}$ for the pressure-entropy formulation, we focus on the formulation with the ``neighbor number density'' volume element used to define $h$, i.e.\ $\tilde{x}_{i}=1$. However, for the sub-sonic, pressure-equilibrium systems simulated in this test, the choice of $\tilde{x}_{i}$ makes little difference (highlighting the importance of our strong shock tests). We also do not further separately consider the pressure-energy formulation, as it gives very similar results in our subsequent tests. As expected from the discussion in \S~\ref{sec:intro}, the density-entropy formulation poorly captures the KH instability. The surface tension term forms a sharp boundary layer between the two phases, which suppresses all the small-wavelength mixing and leads at late times to the breakup of the ``rolls.'' This ultimately produces blobs that resemble an oil-and-water morphology (a system with physical surface tension). 

In the pressure-entropy formulation, however, the behavior is dramatically improved. The long-wavelength mode grows on the correct (linear) KH timescale, but we can also plainly see the growth of modes at all wavelengths (along the ``edge'' of the density surface) down to the kernel scale. At later times, we resolve $\sim3-5$ full ``wraps'' of the rolls while the standard SPH solution breaks up. Most important, there is obviously actual mixing occurring ``inside'' the rolls. Direct comparison to high-resolution results from grid codes (both fixed-grid and adaptive mesh solutions) as well as Godunov SPH and moving-mesh methods, using identical initial conditions, show that the result here is quite similar (compare e.g.\ \citealt{read:2010.sph.mixing.optimization} Fig.~6, or \citealt{murante:2011.godunov.sph} Figs.~5-9).


\begin{figure}
    \plotone{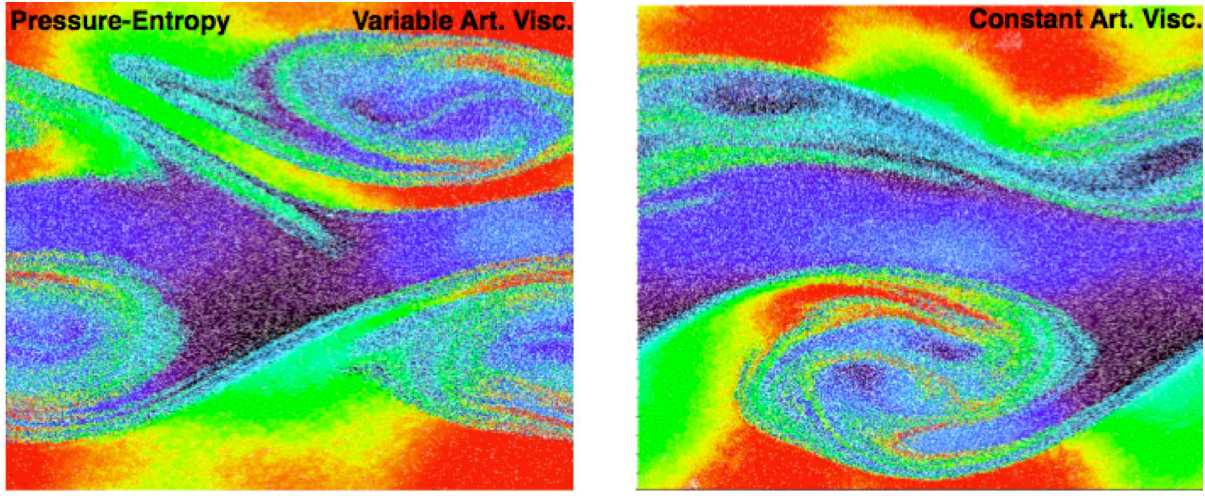}
    \caption{Comparison of the Kelvin-Helmholtz behavior at $t=12\,\tau_{\rm KH}$ on artificial viscosity in the Pressure-Entropy formulation. {\em Left:} Our standard, by-particle time-dependent artificial viscosity following \citealt{morris:1997.sph.viscosity.switch}. {\em Right:} The original {\small GADGET} simplified constant artificial viscosity prescription (as \citealt{gingold.monaghan:1983.artificial.viscosity} with a \citealt{balsara:1989.art.visc.switch} switch). Although very important for some behaviors, it has little effect on KH instabilities within this formulation.
    \label{fig:kh.vs.visc}}
\end{figure}

In Fig.~\ref{fig:kh.vs.visc}, we repeat this experiment with the pressure-entropy formulation, but replace our standard treatment of artificial viscosity (see \S~\ref{sec:sims:viscosity}) with a more simplified treatment (using a constant artificial viscosity for all particles and times); the latter is known to produce significantly more numerical dissipation away from shocks. Although it is well-known that this can produce significant differences in some situations (e.g.\ sub-sonic turbulence and rotating shear flows; see \citealt{cullen:2010.inviscid.sph,price:2010.grid.sph.compare.turbulence,bauer:2011.sph.vs.arepo.shocks}), it has little effect on this particular test. This owes in part to the implementation of the \citet{balsara:1989.art.visc.switch} switch in both cases, which reduces the artificial shear viscosity; without this, we see significantly greater damping of shear motions.

\begin{figure}
    \plotone{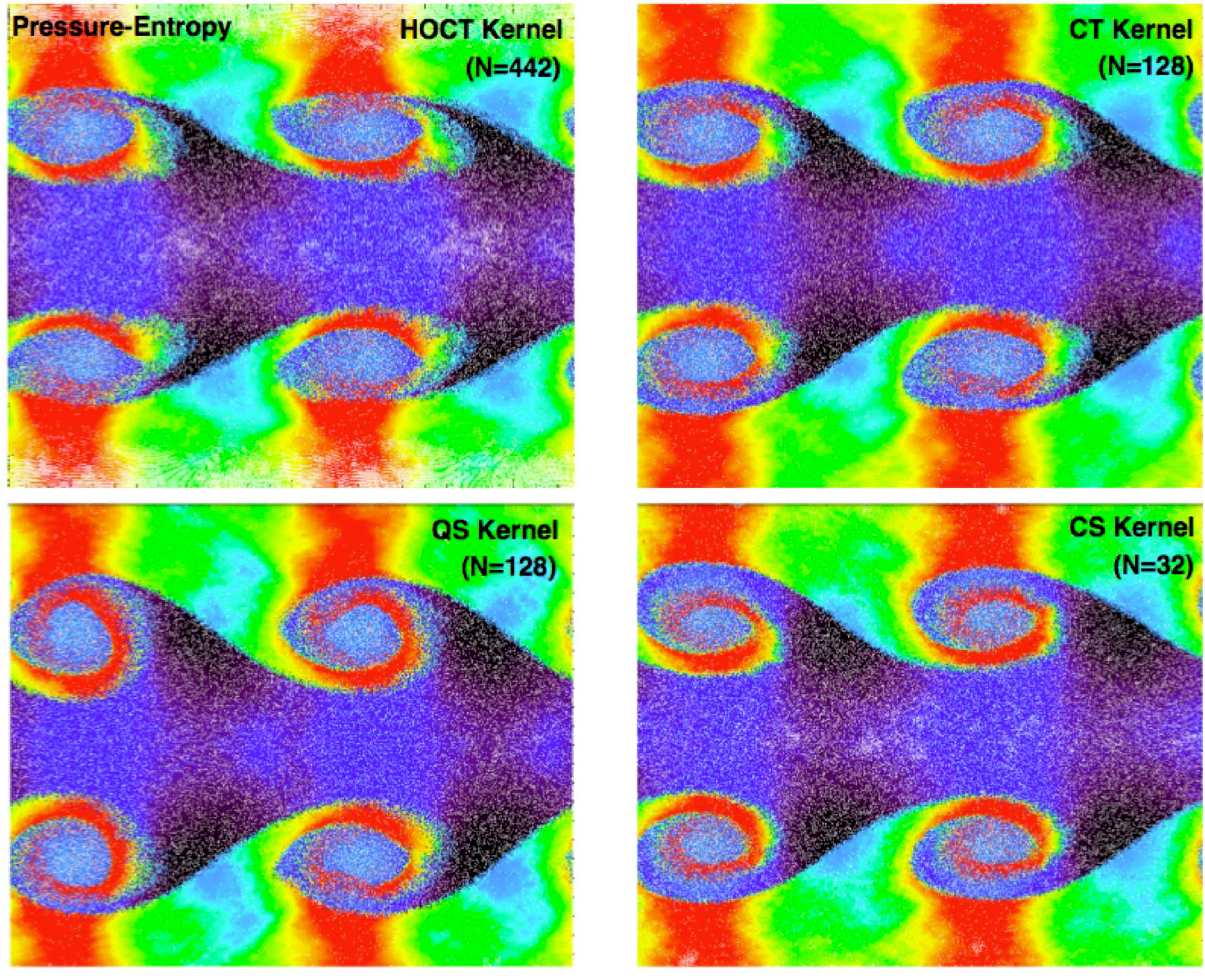}
    \caption{Comparison of the Kelvin-Helmholtz behavior at $t=3\,\tau_{\rm KH}$ on the smoothing kernel in the Pressure-Entropy formulation. {\em Top Left:} Quartic core-triangle kernel with neighbor number $\NNb=442$ from \citet{read:2010.sph.mixing.optimization}. {\em Top Right:} Cubic core-triangle with $\NNb=128$. {\em Bottom Left:} Quintic spline with $\NNb=128$. {\em Bottom Right:} Cubic spline with $\NNb=32$. 
    \label{fig:kh1.vs.kernel}}
\end{figure}

In Fig.~\ref{fig:kh1.vs.kernel}, we again repeat this experiment with the pressure-entropy formulation, but vary the smoothing kernel. We compare two spline kernels, for the neighbor numbers advocated in \citet{price:2012.sph.review,dehnen.aly:2012.sph.kernels}, and the ``core-triangle'' kernels with sharp central peaks, proposed in \citet{read:2010.sph.mixing.optimization}. For the standard initial conditions here, we see only subtle differences over a wide range from even the simplest kernel with $32$ neighbors up through more than an order of magnitude higher neighbor number.\footnote{Note that, at the same error level, the core-triangle kernels require significantly higher neighbor number, because they introduce a sharp bias towards the central particles.} By explicitly eliminating the ``surface tension'' term, while maintaining exact conservation, the pressure-entropy formulation appears to significantly reduce the importance of kernel-level pressure gradient errors in obtaining the correct KH solution (compare e.g.\ the more significant kernel dependence found for standard but non-conservative density-entropy SPH in \citealt{read:2010.sph.mixing.optimization}, Figs.~4 \&\ 6). We have also tested the Wendland kernels proposed in \citet{dehnen.aly:2012.sph.kernels} and find (as they do) essentially identical performance to our spline results for the neighbor numbers here (although the kernels proposed there show greatly improved stability properties at higher neighbor number).

\begin{figure}
   \plotone{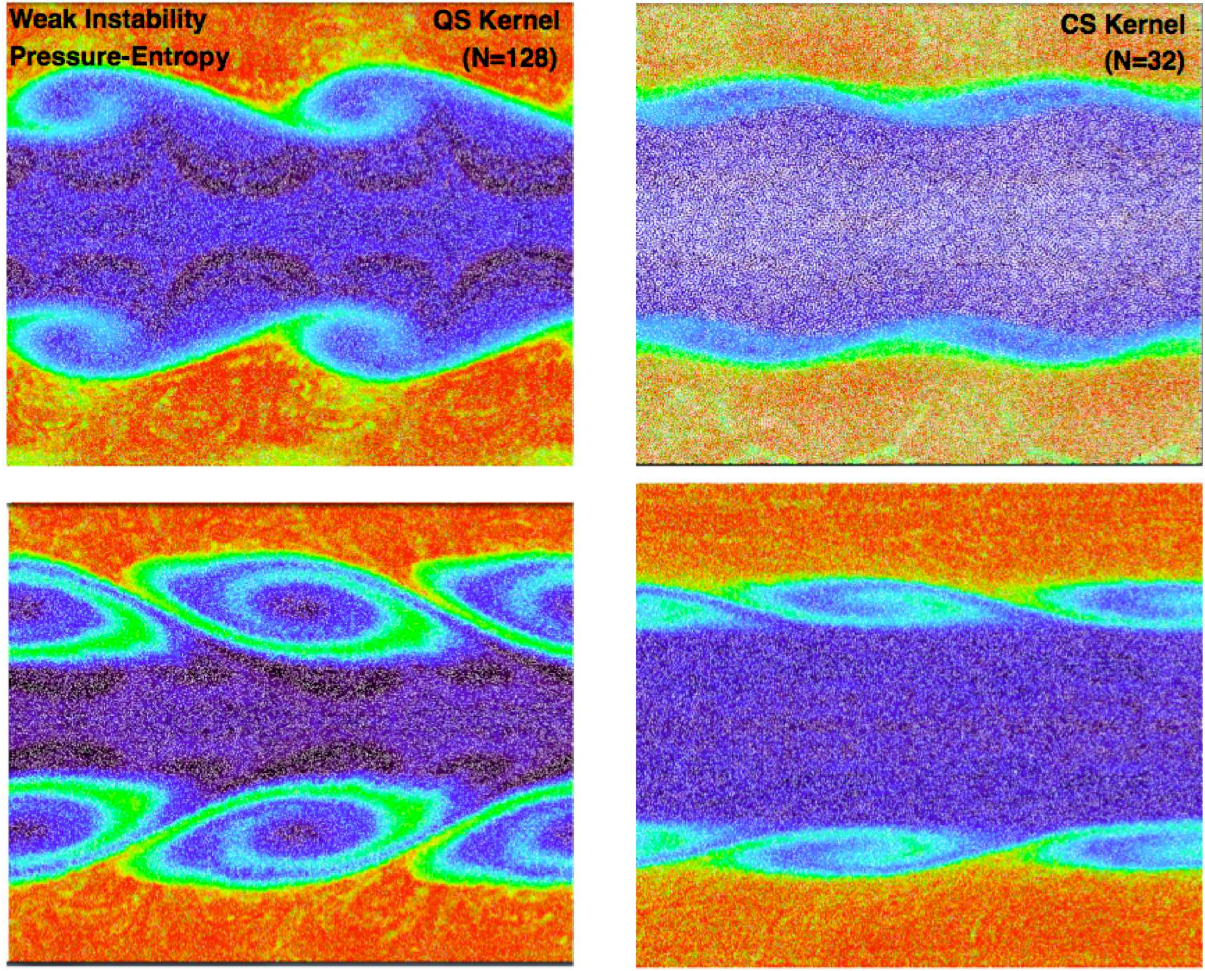}
    \caption{Comparison of Kelvin-Helmholtz behavior with different initial conditions in the Pressure-Entropy formulation at $t\approx1\,\tau_{\rm KH}$ ({\em top}) and $t\approx3\,\tau_{\rm KH}$ ({\em bottom}) on smoothing kernel ({\em left:} quintic spline with $\NNb=128$; {\em right:} cubic spline with $\NNb=32$). Compared to Fig.~\ref{fig:kh1}-\ref{fig:kh1.vs.kernel}, we modify the initial conditions by multiplying the particle number, density, sound speed, pressure, shear velocity, and initial perturbation amplitude by factors of $=0.5,\,2.0,\,2.0,\,8.0,\,0.5,\,0.5$, respectively. All of these changes increase the ratio of particle noise and pressure gradient errors relative to the KH growth. In this limit, the cubic spline with $\NNb=32$, which involves larger gradient errors, is barely able to capture the instability; however the quintic spline with $\NNb=128$ does well. ``Standard'' SPH completely fails to develop any ``curls'' in this limit.
    \label{fig:kh2.vs.kernel}}
\end{figure}

However, in Fig.~\ref{fig:kh2.vs.kernel} we again examine the effects of different kernels (with the pressure-entropy formulation), but with different initial conditions. We reduce the particle number, shear velocity, and initial perturbation amplitude by factors of $2$ and double the initial sound speed; these changes all reduce the magnitude of the initial KH instability and its growth rate, relative to the particle noise and error terms stemming from the kernel sum in the SPH pressure gradients that enter the EOM. This is designed to be challenging (even for grid codes). With this much weaker seeding and noisier particle distribution, we begin to see a dependence on the smoothing kernel. The simplest $\NNb=32$ cubic spline, with the smallest neighbor number, leads to particle noise in the pressure gradients comparable to the actual signal. While the instability is still (barely) captured, the rolls are ``too thin'' and end up being sheared before they reach the proper height and can wrap appropriately. However, our standard quintic spline with $\NNb=128$ performs well, recovering all the key behaviors (note that the non-linear behavior is different here than in the previous test, as it should be for the different pressure and shear velocities). Going to still higher resolution or varying the kernel at higher $\NNb$ gives well-converged results at this point.

\begin{figure}
   \plotone{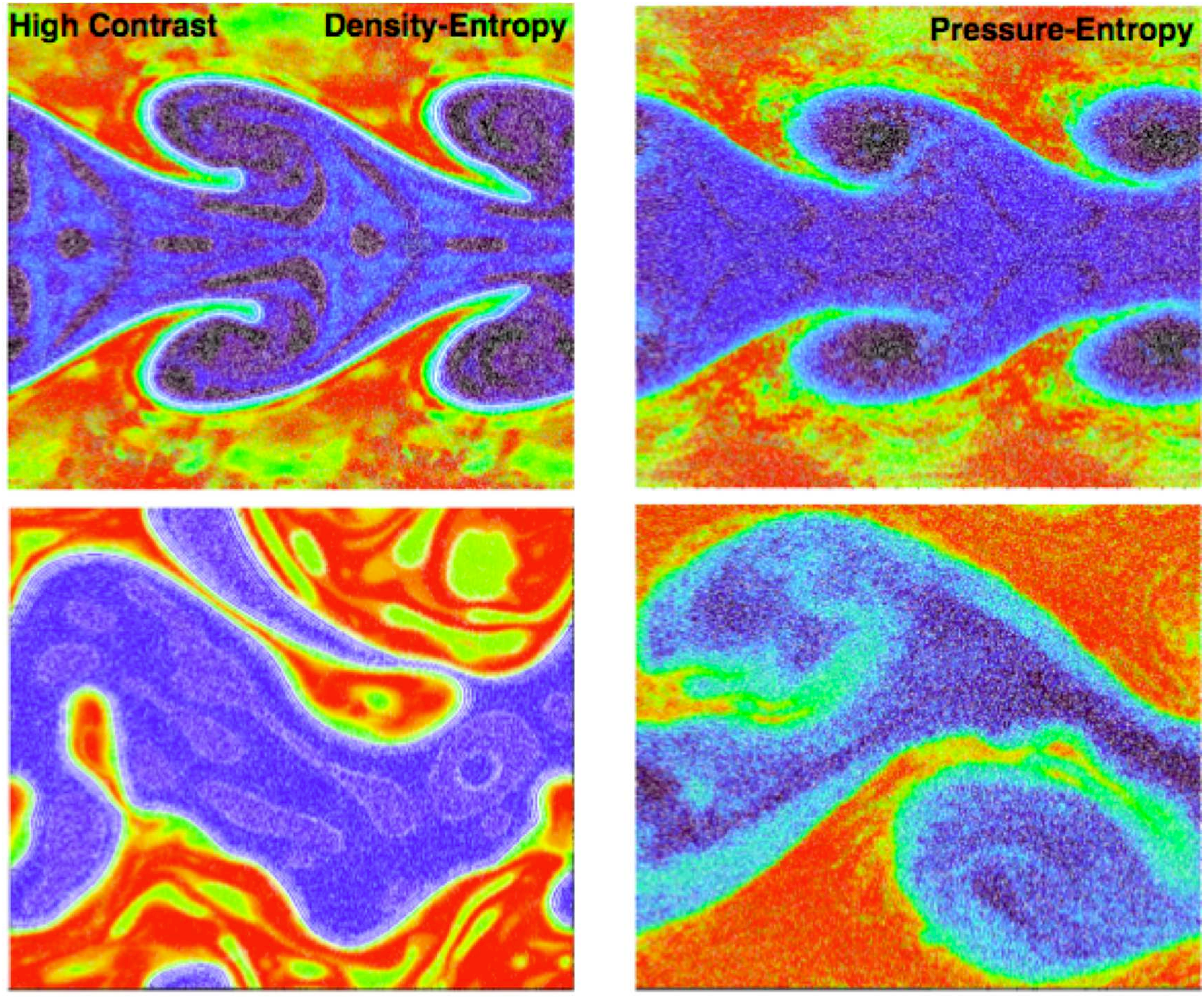}
    \caption{Comparison of KH instabilities at $t\approx2\,\tau_{\rm KH}$ ({\em top}) and $t\approx10\,\tau_{\rm KH}$ ({\em bottom}), for the ``standard'' (density-entropy) SPH ({\em left}) and pressure-entropy ({\em right}; with $\tilde{x}_{i}=1$) SPH formulations. The initial conditions are as Fig.~\ref{fig:kh1}, but the initial density contrast is increased from a factor of $2$ to a factor of $20$, with the initial particles no longer being equal mass but $4$ times more massive in the initial high-density region. Sharp boundary layers that lead to ``gloopy'' morphology are evident in standard SPH. In Pressure-Entropy SPH it remains well-behaved, although hints of ``gloopiness'' appear in the transition between linear growth and fully non-linear instability.
    \label{fig:kh.highcontrast}}
\end{figure}

In Fig.~\ref{fig:kh.highcontrast}, we repeat our KH test again but multiply the initial density contrast by a factor of $10$, and instead of using constant-mass particles we use particles with masses a factor of $4$ larger in the high-density region. As discussed in \citet{read:2012.sph.w.dissipation.switches}, many proposed alternative formulations of SPH, designed to improve fluid mixing, fail in this regime (see e.g.\ Fig~6 in \citealt{read:2010.sph.mixing.optimization} and Figs.~A1 \&\ E1 in \citealt{read:2012.sph.w.dissipation.switches}). And we see an even more pronounced ``boundary layer'' separating the phases in the standard density-entropy SPH formulation. This occurs because the higher density contrast exacerbates any (even small residual) surface tension term, and the multi-mass particles increase particle disorder and leading-order errors in the pressure gradient estimator. Multi-mass particles also make it critical to have a well-behaved criterion for smoothing lengths, and increase the errors from neglecting $\nabla h$ terms. But we see that the pressure-entropy formulation remains well-behaved in this case. There is some increased hint of ``gloopiness'' as the instability transitions between linear and non-linear growth, but at least some of this is because of the (correct) slower growth of the small-wavelength modes. 

Finally, it is worth noting (though perhaps as more of a curiosity) that the ``poor'' solution of the density-entropy formulation in the cases above looks very similar to the ``correct'' solution for a modestly magnetized medium (with some field component parallel to the shear or tangled; see e.g.\ \citealt{frank:1996.mhd.kh} and references therein). This occurs because, if we consider the linear perturbation analysis of the KH instability, the ``incorrect'' surface tension force here is (for the initial linear stage) almost mathematically identical to a ``correct'' magnetic tension term for parallel fields with strength $\beta\sim1$.

\vspace{-0.5cm}
\subsection{Rayleigh-Taylor Instabilities}

We now consider the Rayleigh-Taylor (RT) instability, with initial conditions from \citet{abel:2011.sph.pressure.gradient.est}. In a two-dimensional slice with $512^{2}$ particles and $0<x<1/2$ (periodic boundaries) and $0<y<1$ (reflecting boundary with particles at initial $y<0.1$ or $y>0.9$ held fixed), we take $\gamma=1.4$ and initialize a density profile $\rho(y)=\rho_{1}+(\rho_{2}-\rho_{1})/(1+\exp{[-(y-0.5)/\Delta]})$ where $\rho_{1}=1$ and $\rho_{2}=2$ are the density ``below'' and ``above'' the contact discontinuity and $\Delta=0.025$ is its width; initial entropies are assigned so the pressure gradient is in hydrostatic equilibrium with a uniform gravitational acceleration $g=-1/2$ in the $y$ direction (at the interface, $P=\rho_{2}/\gamma=10/7$ so $c_{s}=1$). An initial $y$-velocity perturbation $v_{y} = \delta v_{y}\,(1+\cos{(8\pi\,(x+1/4))})\,(1+\cos{(5\pi\,(y-1/2))})$ with $\delta v_{y}=0.025$ is applied in the range $0.3<y<0.7$.

\begin{figure*}
    \plotsidesize{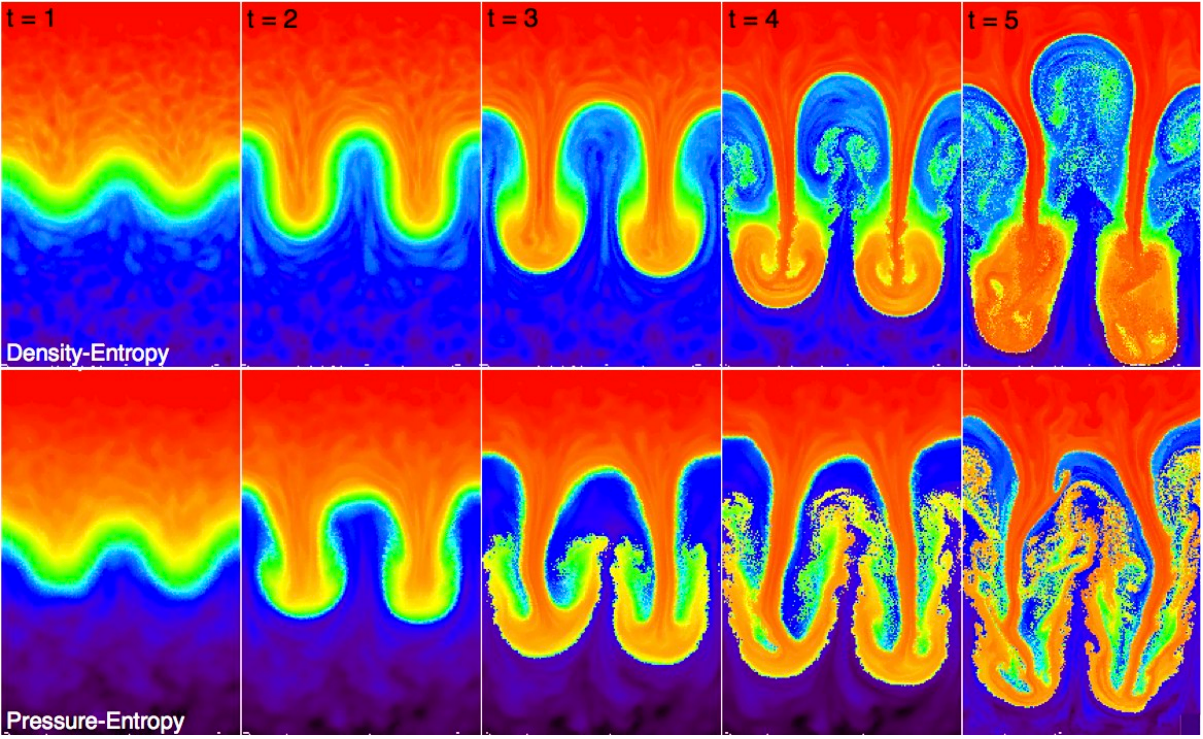}{0.9}
    \caption{Time evolution of a two-dimensional Rayleigh-Taylor (RT) instability (specific entropy map shown). {\em Top:} Standard (density-entropy) SPH. {\em Bottom:} Pressure-entropy formulation ($\tilde{x}_{i}=1$). The RT instability develops in both cases but the mixing along the rising/sinking surfaces (a KH instability) is suppressed in density-entropy SPH, leading to different nonlinear outcomes. 
    \label{fig:rt}}
\end{figure*}

Fig.~\ref{fig:rt} shows the resulting evolution as a function of time, in the density-entropy and pressure-entropy formulations. As in \SM, both cases develop the RT instability with a similar linear growth time (slightly slower in the density-entropy case); we find that this is true for all of the kernel variations and both artificial viscosity choices discussed above, as well as for the slightly different (isoentropic) initial conditions used in \SM, also for $10$-times smaller initial perturbations ($\delta v_{y}=0.0025$), and for resolutions as low as $50x100$ particles.\footnote{This agrees with the results in \SM, as well as the development of RT instabilities in blastwaves seen in other density-entropy implementations \citep[see e.g.][]{herant:1994.sph.tricks,price:2011.sph.turb.notes}. It is not entirely clear why the RT instability fails to develop in standard SPH with the same initial conditions in \citet{abel:2011.sph.pressure.gradient.est}, but there are a number of additional differences in the algorithms (see \S~\ref{sec:sims}). It may relate to the fact that the kernel instabilities discussed above can be much more severe in 2D standard SPH, requiring careful matching of neighbor number and kernel shape.} However, they differ significantly in their nonlinear evolution; surface tension in the density-entropy formulation prevents the development of fine structure in the shear flow along the ``fingers,'' obviously very closely related to the behavior in the KH tests. Pressure-entropy SPH, however, exhibits growth on the correct linear timescale and nonlinear behavior in agreement with that in Eulerian and moving-mesh schemes \citep[e.g.][]{springel:arepo}.

\vspace{-0.5cm}
\subsection{The ``Blob'' Test}

We next consider the ``blob'' test, which is designed to test processes of astrophysical interest (e.g.\ ram-pressure stripping, mixing, KH and Rayleigh-Taylor instabilities in a multiphase medium). Again our initial conditions come from the Wengen test suite and are described in \citet{agertz:2007.sph.grid.mixing}. Briefly, the initial conditions include a spherical cloud of gas with uniform density in pressure equilibrium with an ambient medium, in a wind-tunnel with periodic boundary conditions. The imposed wind has Mach number $\mathcal{M}=2.7$ and the initial density/temperature ratios are $=10$. The box is a periodic rectangle with dimensions $x,\,y,\,z=2000,\,2000,\,6000\,$kpc, with the cloud centered on $0,\,0,\,-2000\,$kpc; the $9.6\times10^{6}$ particles are equal-mass and placed on a lattice. 

\begin{figure}
\begin{tabular}{c}
    \includegraphics[width={0.95\columnwidth}]{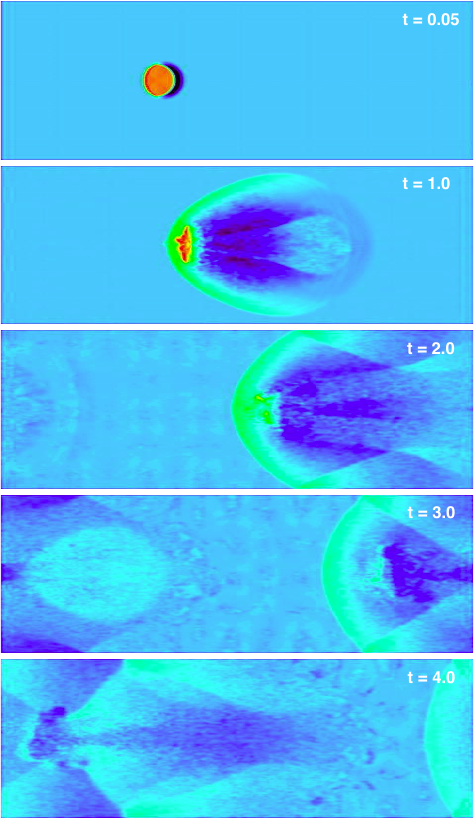} 
\end{tabular}
    \caption{Central density slices in the ``blob test'' (a high-density, pressure-equilibrium cloud hit by a wind) with the pressure-entropy formulation. Time (in units of the KH growth time) increases from top to bottom. The high-density (red/orange) cloud gas is efficiently mixed by instabilities within a couple cloud crossing times; the morphology and density distribution agree well with grid codes.
    \label{fig:blob}}
\end{figure}

\begin{figure}
    \plotone{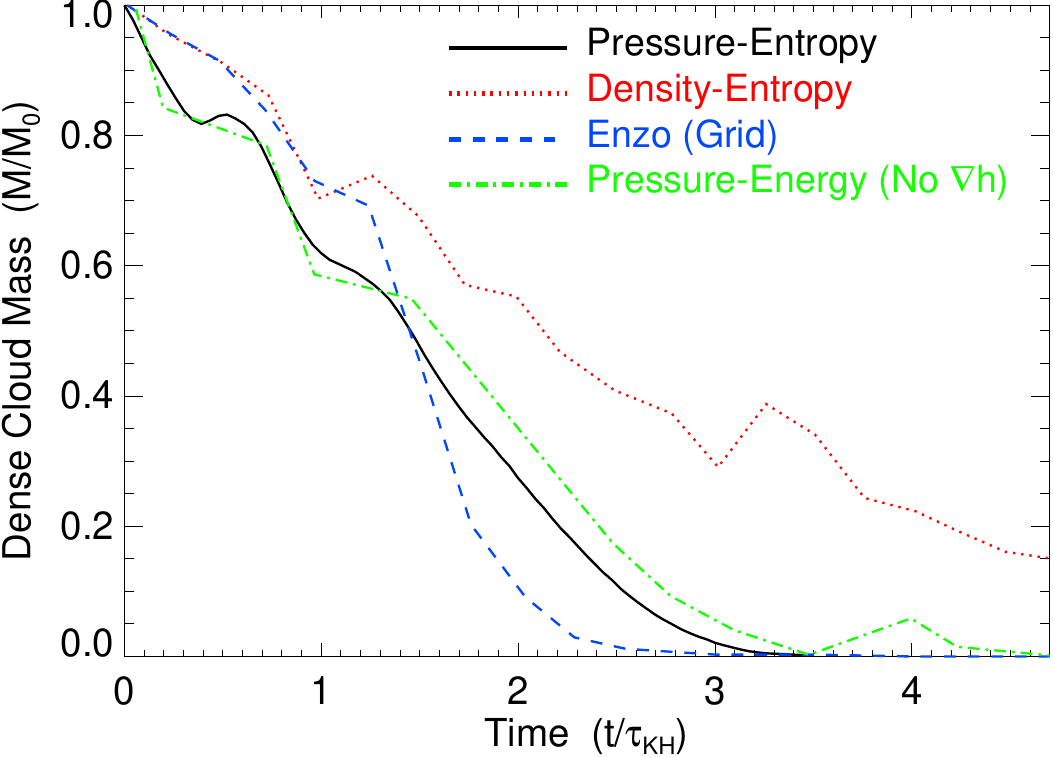} \\
    \caption{Fraction of the initial cloud in Fig.~\ref{fig:blob} which remains both cold and dense (i.e.\ avoids mixing) as a function of time, relative to the KH growth time at the cloud surface. We compare the standard SPH (density-entropy) and pressure-entropy formulations here, as well as the pressure-energy formulation in \SM\ (which does not include the $\nabla h$ terms) and the results from a high-resolution grid code method (Enzo). The grid code and pressure formulations (independent of the $\nabla h$ terms) agree reasonably well. Density formulations show slower mixing/stripping. 
    \label{fig:blob.decay}}
\end{figure}

Fig.~\ref{fig:blob} shows the resulting morphology of the cloud as a function of time, in the pressure-entropy formulation. Fig.~\ref{fig:blob.decay} attempts to quantify the rate of cloud disruption following \citet{agertz:2007.sph.grid.mixing}, who defined a useful standard criterion for measuring the degree of mixing of initially dense blob gas: at each time, we simply measure the total mass in gas with $\rho>0.64\,\rho_{c}$ and $T<0.9\,T_{a}$ (where $\rho_{c}$ and $T_{a}$ are the initial cloud density and ambient temperature). We show the standard SPH (density-entropy) result from that paper, as compared to the prediction from the pressure-entropy formulation here. We also compare the result from the identical test in \SM, using the pressure-energy formulation but without including the $\nabla h$ terms in the EOM, as well as the results from a high-resolution run with Enzo \citep{oshea:2004.enzo.introduction}, a grid code. 

The wind-cloud interaction generates a bow shock and immediately begins disrupting the cloud via a combination of KH and RT instabilities at its surface. By the middle panel in Fig.~\ref{fig:blob}, there is no visible large concentration of dense material. Compare this to the identical initial conditions run with standard SPH and grid codes in \citet{agertz:2007.sph.grid.mixing} (their Figs.~4 \&\ 7). In standard SPH, the cloud is compressed to a ``pancake,'' but the tension term prevents mixing at the surface and so a sizeable fraction survives disruption for large timescales. In contrast, the predicted morphology of the cloud here agrees very well with that in adaptive mesh codes therein (as well as moving-mesh methods in \citealt{sijacki:2011.gadget.arepo.hydro.tests}). And quantitatively, we see this in Fig.~\ref{fig:blob.decay}. 

However, once again we should note that the solution derived from the density-entropy formulation is remarkably similar to the ``correct'' solution with magnetic fields with $\beta\gtrsim1$ (compare e.g.\ \citealt{maclow:mhd.shock.cloud,shinstone:3d.mhd.shock.cloud}), because the artificial surface tension term acts similarly to magnetic tension.

\begin{figure}
\begin{tabular}{c}
    \includegraphics[width={0.95\columnwidth}]{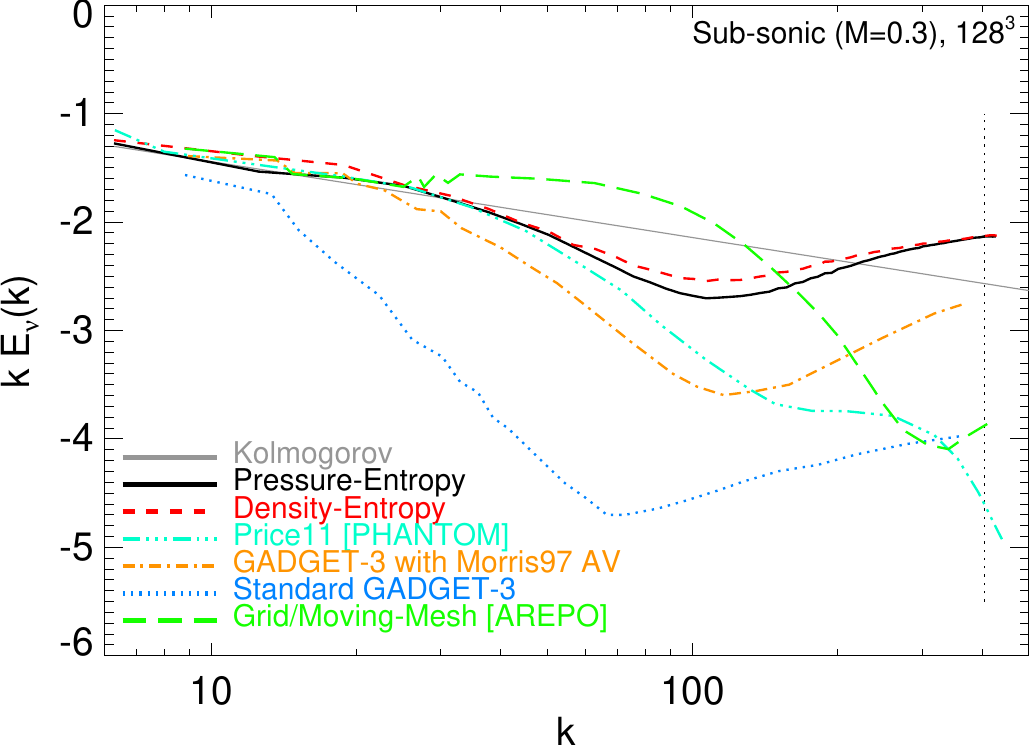} 
\end{tabular}
    \caption{
    Velocity power spectrum in sub-sonic ($\mathcal{M}=0.3$) driven, isothermal turbulence. Each simulation uses $128^{3}$ particles and identical driving. We compare the analytic Kolmogorov inertial-range model, to that calculated from our standard density-entropy (Eq.~\ref{eqn:eom.density.entropy}) and pressure-entropy (Eq.~\ref{eqn:eom.pressure.entropy.betterh}) formulations. The mean softening length $h$ is shown as the vertical dotted line. Both do well down to a few softenings, where they first fall below the analytic result (excess dissipation) then rise above (kernel-scale noise). The EOM choice has a weak effect on the results. We compare different SPH algorithms and codes (description in text); agreement is good where the same methods are used. The ``{\small PHANTOM}'' and ``{\small GADGET-3} with Morris97 AV'' use a density-entropy EOM with similar artificial viscosity treatment to our calculations (which dominates the inertial range), but different SPH kernels and power spectrum calculation methods (which dominate the noise at $\lesssim 3\,$h). The ``standard'' {\small GADGET-3} calculation uses a constant artificial viscosity and different kernel, and produces almost no inertial range. The {\small AREPO} calculation considers a grid/moving mesh at the same resolution: deviations from Kolmogorov occur around the resolution limit, but are of a different character. 
    \label{fig:subsonic}}
\end{figure}

\vspace{-0.5cm}
\subsection{Sub-Sonic Turbulence}

\citet{bauer:2011.sph.vs.arepo.shocks} present a detailed study of the properties of idealized driven isothermal turbulence in SPH -- specifically using {\small GADGET-3} with the density-entropy formulation -- as compared to moving mesh and grid methods (in the code {\small AREPO}; \citealt{springel:arepo}). Consistent with earlier results, they found that different methods agree well when turbulence is super-sonic \citep[e.g.][]{kitsionas:2009.grid.sph.compare.turbulence,price:2010.grid.sph.compare.turbulence}. But when the turbulence is sub-sonic, they found that density-entropy SPH tended to reproduce a smaller inertial range. However as discussed there and in \citet{price:2011.sph.turb.response}, this can depend quite sensitively on the artificial viscosity prescriptions and other numerical details. We therefore reproduce the sub-sonic, driven turbulence experiment from \citet{bauer:2011.sph.vs.arepo.shocks} but adopt the pressure-entropy formulation, to test whether it also differs significantly from the results in the density-entropy formulation. We adopt the identical setup and resolution (A.\ Bauer, private communication), with the initial conditions and driving algorithm described therein. Briefly, we initialize a box of unit length, density, and sound speed with $128^{3}$ particles, and drive the turbulence in a narrow range of large-$k$ modes with characteristic Mach number $\mathcal{M}=0.3$ on the largest scales; unlike our other experiments, the gas is isothermal ($\gamma=1$). We run the experiment to $t=25$ (more than sufficient to reach steady-state).


We compare the turbulent velocity spectrum measured following \citet{bauer:2011.sph.vs.arepo.shocks} (we linearly interpolate the particle-centered velocity values to a uniform grid, fourier transform each velocity component, and average in bins of fixed $|{\bf k}|$ to obtain the kinetic energy power spectrum). The resulting power spectrum is plotted from $k\approx2\pi$ (though $k\lesssim10$ are affected by the turbulent forcing as well as the finite box size) to $k\approx500$ (a smoothing length $h$). We compare the pressure-entropy formulation to the results of standard (density-entropy formulation) SPH. The results are nearly identical (although there is slightly more power at intermediate scales in the density-entropy formulation). The results follow the expected Kolmogorov inertial range, until $\sim4\,h$, where they drop below owing to artificial numerical dissipation; they then rise at $\sim2\,$h, owing to kernel-scale noise (largely from pressure gradient errors). 

We can compare this to other SPH results using the identical initial conditions and driving routine. First, consider the results from \citet{bauer:2011.sph.vs.arepo.shocks}, using ``standard'' {\small GADGET-3} (density-entropy EOM, without the additional algorithm improvements discussed in \S~\ref{sec:sims}) but with an artificial viscosity scheme following \citet{morris:1997.sph.viscosity.switch} similar (though slightly different) to that adopted here, and the noisier $\NNb=32$ cubic spline kernel. We also compare the results from \citet{price:2011.sph.turb.response} using a different SPH code ({\small PHANTOM}), with the same density-entropy EOM, and an artificial viscosity treatment identical to that here, but using an intermediate kernel ($\NNb=58$ cubic spline) and different power spectrum calculation method. As shown there, the artificial viscosity treatment dominates large/intermediate scales. So with identical artificial viscosity, the \citet{price:2011.sph.turb.response} result is identical to ours over the inertial range (including the initial ``drop off''). The smaller-scale ($\lesssim2-3\,h$) behavior is dominated by the kernel choice and method of power spectrum calculation. If we compare ``standard'' {\small GADGET-3} with constant artificial viscosity \citep{gingold.monaghan:1983.artificial.viscosity} from \citet{bauer:2011.sph.vs.arepo.shocks}, we see much more severe excess dissipation, and almost no inertial range. 

Comparing to the results from {\small AREPO} in \citet{bauer:2011.sph.vs.arepo.shocks} (run either in fixed-grid or moving-mesh mode, they are nearly identical), we see that the deviations from Kolmogorov there are opposite in sign and quite distinct. In SPH, artificial viscosity produces excess dissipation on larger (resolved) scales, while kernel gradient errors lead to particle noise that boosts the power at the smallest scales. The former (larger-scale deviations), being dominated by artificial viscosity, are only indirectly tied to resolution; considering higher-resolution runs we confirm the results in \citet{bauer:2011.sph.vs.arepo.shocks} regarding the relatively slow convergence the high-$k$ spectrum in SPH relative to grid codes. The latter (smaller-scale deviations) are related to the conservative nature of the code and its ability to dissipate particle noise (their ``return'' to and overshoot of the Kolmogorov power-law are numerical, rather than physical consequences of a turbulent cascade). In Eulerian approaches, the lack of any but numerical viscosity concentrates the dissipation at the resolution/cell scale, which produces a ``bottleneck'' of excess power that cannot be dissipated at the larger (marginally resolved) scales; but this scales directly with the resolution.

\begin{figure}
    \plotone{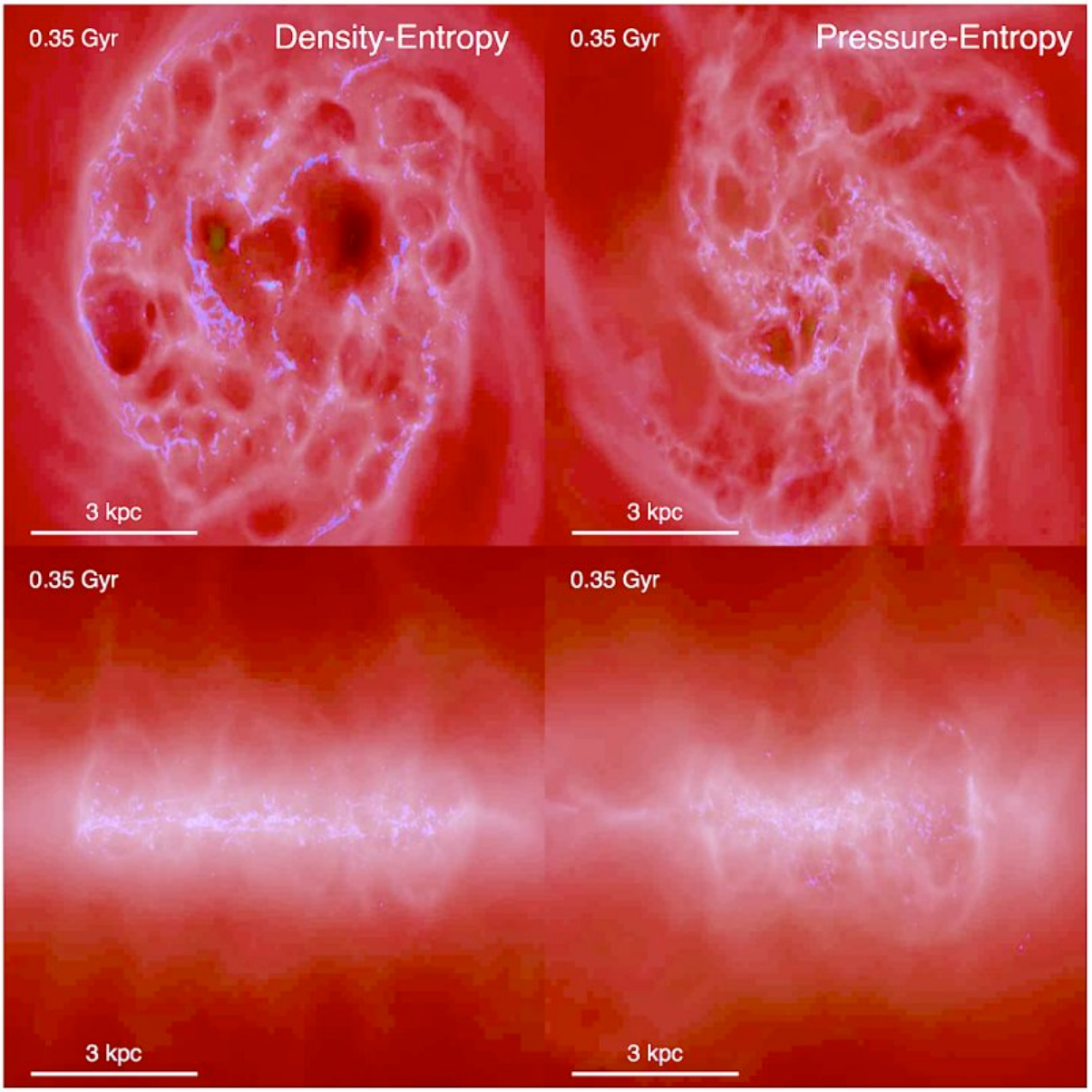} 
    \caption{Morphology of the gas in a simulation of a star-forming disk galaxy (an isolated, SMC-mass dwarf) from \citet{hopkins:fb.ism.prop}, following the resolved physics of the ISM and stellar feedback explicitly. Brightness encodes projected gas density (logarithmically scaled with $\approx6\,$dex stretch); color encodes gas temperature with the blue material being $T\lesssim1000\,$K molecular and atomic gas, pink $\sim10^{4}-10^{5}$\,K warm ionized gas, and yellow $\gtrsim10^{6}\,$K hot gas. We show the galaxy face-on {\em top} and edge-on {\em bottom} after several orbital periods of evolution (during which the galaxy is in quasi-steady state). We compare the identical simulation using the density-entropy (Eq.~\ref{eqn:eom.density.entropy}) and pressure-entropy (Eq.~\ref{eqn:eom.pressure.entropy.betterh}) formulations of SPH. The two are largely similar. There is slightly sharper delineation between phases (e.g.\ molecular clouds and hot bubbles) in the density-entropy formulation; this includes the sharper transition between the star-forming and outer disks (physically, where the cooling time becomes longer than the dynamical time). 
    \label{fig:smc.morph}}
\end{figure}
\begin{figure}
    \plotone{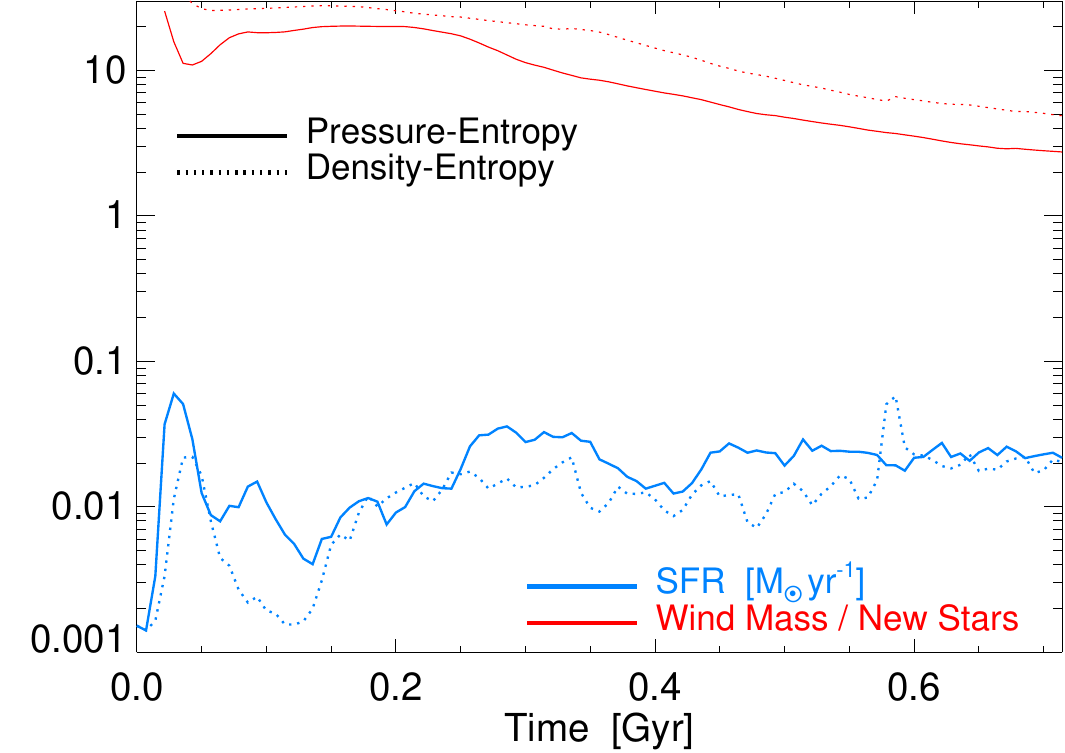} 
    \caption{Star formation rate and stellar wind mass-loading (ratio of total unbound gas mass to total new stellar mass formed since the start of the simulation) as a function of time, for the galaxy simulations in Fig.~\ref{fig:smc.morph}. Although stochastic variations in the SFR differ by factors $\sim2-3$, the time-averaged SFR is only $\sim20\%$ lower in the density-entropy formulation. The wind mass-loading is systematically higher in this case by a factor $\sim1.5-2.0$. This stems from increased phase mixing in the pressure-entropy formulation introducing more cold, dense gas into ``hot'' gas, enhancing its cooling before it can vent out of the disk.
    \label{fig:smc.sfr}}
\end{figure}

\vspace{-0.5cm}
\subsection{Stellar Feedback in Isolated Galaxies}
\label{sec:feedback.example}

Finally we test our modified algorithm on a fully non-linear system of particular astrophysical interest. Specifically, we re-run one of the star-forming galaxy disk models described in \citet{hopkins:rad.pressure.sf.fb} and a subsequent series of papers \citep{hopkins:fb.ism.prop,hopkins:stellar.fb.winds,hopkins:clumpy.disk.evol}. These simulations include gravity, collisionless stars and dark matter, and gas with a wide range of cooling processes\footnote{Cooling requires a density estimate, independent of whether one is needed for the EOM. For this, we use the standard SPH kernel density estimator, for the reasons in \S~\ref{sec:density.estimator}.} from $\sim10-10^{9}\,$K, super-sonic turbulence, shocks, star formation in resolved giant molecular clouds, and explicit treatment of stellar feedback from SNe Types I \&\ II, stellar winds, photoionization and photoelectric heating, and radiation pressure. Our purpose here is both to test how robust these algorithms are (since many problems with e.g.\ non-linear error terms or conservation errors often do not manifest in simple test problems such as those above), and to examine how significant the pure numerical issues of fluid mixing are, relative to e.g.\ the previously-studied effects of adding/removing different physics in these simulations. 

We specifically re-run the ``SMC'' model (a model of an Small Magellanic Cloud-mass, gas-rich isolated dwarf galaxy), at ``high'' resolution ($1\,$pc softening length) as defined in the papers above, with all the above physical processes included exactly as in \citet{hopkins:fb.ism.prop}, but in one case adopting the density-entropy formulation and in another the pressure-entropy formulation. We choose this model because, of the galaxy types studied therein, it has the largest mass fraction in hot gas and is most strongly affected by thermal feedback mechanisms (e.g.\ SNe), and as a result is most sensitive to fluid mixing and phase structure. Figs.~\ref{fig:smc.morph}-\ref{fig:smc.sfr} compare the visual morphologies and star formation rates from these simulations. Taking into account that the systems are turbulent and non-linear, the morphologies are similar. They differ as one might expect: the pressure-entropy formulation increases mixing along phase boundaries, leading to less-sharp divisions between molecular clouds and/or hot, under-dense bubbles and the surrounding medium. This also makes the division between the star-forming and ``extended'' disk less sharp, as it is largely determined by the radius where the cooling rate becomes fast relative to the dynamical time.\footnote{We have also attempted a comparison using the algorithm in \citet{abel:2011.sph.pressure.gradient.est}, from Fig.~\ref{fig:sedov2}. Over short timescales this gives similar results to the pressure-entropy formulation. However, we cannot evolve the runs very long before a situation similar to Fig.~\ref{fig:sedov2} arises when, for example, a SNe occurs in an under-dense region and its energy is deposited in a small number of particles, leading to large conservation errors.}

Quantitatively, the time-averaged star formation rates differ remarkably little, just $\approx20\%$ (with higher SFRs in the pressure-entropy case); this owes to the fact that they are globally set by equilibrium between momentum injection from stellar feedback and dissipation of turbulent energy/momentum on a dynamical time \citep{hopkins:fb.ism.prop}. The instantaneous rates vary more rapidly and can differ by factors of $\sim 2-3$. We also compare the mass-loading of galactic winds in each simulation, defined as in \citet{hopkins:stellar.fb.winds} as the ratio of total ``wind mass'' (material with positive Bernoulli parameter, i.e.\ which will be unbound in the absence of external pressure forces) to new stellar mass formed since the beginning of the simulation. Here the qualitative behavior is the same but the wind mass-loading is systematically smaller in the pressure-entropy case, by a factor of $\sim1.5-2$. This is directly related to the higher SFR; both owe to the increased mixing reducing the cooling time of the hot gas bubbles (which provide, for this small galaxy, much of the wind mass). In more massive galaxies, the winds in these simulations are less dominated by hot gas, so the effect will be smaller. These differences (while not negligible) are smaller than those typically caused by adding or removing different feedback mechanisms, or even changing details of their implementation \citep[shown in][]{hopkins:stellar.fb.winds}. For example, removing stellar feedback entirely in this simulation changes the SFR by a factor of $\sim50$! Integrated over cosmological times, or allowing for different galaxy conditions because of the integrated effects of different heating/cooling efficiencies as gas actually accretes into halos, however, these small differences can easily build up into more significant divergence \citep[see e.g.][]{vogelsberger:2011.arepo.vs.gadget.cosmo}.

\vspace{-0.5cm}
\section{Discussion}
\label{sec:discussion}

With inspiration from \SM, we re-derive a fully self-consistent set of SPH equations of motion which are independent of the kernel-calculated density, and therefore remove the well-known ``surface tension'' terms that can suppress fluid mixing. The equations still depend on the medium having a differentiable pressure (hence still require some artificial viscosity to capture shocks), but unlike the traditional SPH EOM, remain valid through contact discontinuities. 

Our derivation of the EOM relies on the key conceptual point in \SM, that the SPH volume element does not have to explicitly involve the mass density. However, our derivation and resulting EOM adds four key improvements: {\bf (1)} We rigorously derive the equations from the discretized particle Lagrangian. This guarantees one of the most powerful features of SPH, namely {\em manifest} simultaneous conservation of energy, entropy, momentum, and angular momentum, and an exact solution to the particle continuity equation. {\bf (2)} We similarly derive the ``$\nabla h$'' terms, which are {\em required} for manifest conservation if the SPH smoothing lengths $h_{i}$ are not everywhere constant. {\bf (3)} We derive an ``entropy formulation'' of the equations that allows for the direct evolution of the entropy, avoiding the need to construct/evolve an energy equation, and gives better entropy conservation properties as in \Sentropy; this also happens to minimize the correction terms involved in using a ``particle neighbor number'' definition to define $h$, as compared to the ``energy formulation.'' {\bf (4)} We show how the Lagrangian derivation can be generalized to {\em separate} definitions of the thermodynamic volume element (relating e.g.\ $P$ and $u$) and that used to define the smoothing lengths. This resolves problems of numerical stability and excess diffusion in strong shocks and/or large density contrasts, and automatically allows for varying particle masses.

In fact, we derive a completely general, Lagrangian form of the EOM, including the $\nabla h$ terms, for {\em any} definition of the SPH thermodynamic volume element. Essentially {\em any} particle-carried quantity can be used in the kernel sum entering the EOM, and any (not necessarily the same) differentiable function used to define how the smoothing lengths $h_{i}$ are scaled. In some ways this replaces the long-known ``free weighting functions'' used to define the SPH EOM in their original ``discretized volume element'' formulation. However, in that approach, the choice of different functions generically violates conservation and continuity; here, we demonstrate that a similar physical degree of freedom can be utilized in the discretization of the equations of motion without such violations. 

Based on this degree of freedom, it is easy to see how different discretizations of the EOM might be optimized for some problems. By choosing the required kernel-evaluated element to directly represent a very smooth/stable property in the system, one not only removes spurious ``tension'' terms associated with discontinuities in other system variables, but also minimizes the inevitable discretization error from representing these quantities with a kernel sum. For the constant-pressure (but mixed density) fluid mixing tests we show here, the optimal choice is the ``pressure formulation.'' In MHD applications this kernel sum could trivially be altered to include the magnetic pressure. However if simulating an incompressible or weakly compressible fluid, the ``density formulation'' may well be superior. Direct kernel sums of nominally ``higher order'' properties such as the vorticity or vortensity are also valid and may represent useful formulations for some problems. It is even possible (in principle) to generalize our derivation to one in which different particle subsets have differently-defined volume elements; although we caution that such an approach requires great care.  

For the test problems here, we show that the ``pressure-entropy'' and ``pressure-energy'' formulations  dramatically improve the treatment of fluid interface instabilities including the Kelvin-Helmholtz instability, Rayleigh-Taylor instability, and the ``blob test'' (a mix of Kelvin-Helmholtz and Raleigh-Taylor instabilities as well including non-linear evolution and shock capturing); giving results very similar to grid methods. They also remove the ``deforming'' effect of the surface tension term (allowing, for example, the long-term evolution of an irregular shape of gas at constant pressure but high density contrast); deformation is difficult to avoid even in grid codes (unless the chosen geometry matches the grid), and would otherwise require moving-mesh approaches to follow. However, unlike some of the modifications in the literature proposed to improve the fluid mixing in SPH (which violate conservation), the manifest conservation properties of our derivation mean that it remains well-behaved even in very strong shocks and does not encounter problems of either energy conservation or particle order in e.g.\ extremely strong blastwave problems. 


With these changes in place, we find weaker (albeit still significant) residual effects from  improvement in the artificial viscosity scheme. Comparisons of such schemes are well-studied and what we implement here is still not the most sophisticated possible treatment, although it still considerably reduces artificial viscosity away from shocks \citep[for more detailed studies, see e.g.][]{cullen:2010.inviscid.sph}. We find similar effects from changes to the SPH smoothing kernel. Our favored kernel is taken from more detailed kernel comparison studies in \citet{hongbin.xin:05.sph.kernels,dehnen.aly:2012.sph.kernels}; however, unlike some other SPH formulations, we find that even the ``simplest'' kernel possible ($\NNb=32$ cubic spline) reproduces good results in several tests, except in the expected regime where we wish to resolve kernel-scale growing instabilities that rely on sub-sonic motions at the level of $\mathcal{M} \lesssim N_{\rm NGB}^{-1}$ and so summation errors dominate. This relates to the manifest conservation and maintenance of good particle order implicit in the EOM \citep[see][]{price:2012.sph.review}.

We test the algorithm not just in the ``standard'' set of test problems but also an example of direct astrophysical interest, simulating the evolution of galaxies with a multi-phase ISM. This is useful because it makes clear that for this problem, at least, the differences arising from the treatment of different physics (e.g.\ how cooling, star formation, stellar feedback, and AGN feedback are implemented) makes, on average, larger differences than the numerical scheme \citep[also shown in other code comparisons; e.g.][]{scannapieco:2012.aquila.cosmo.sim.compare}. This is not surprising: those choices lead to orders-of-magnitude differences as opposed to the (still significant) factor $\sim$couple effects of numerical choices. Moreover the differences we are concerned with here largely pertain to mixing in sub-sonic, non-radiative flows dominated by thermal pressure; in contrast many astrophysical problems of interest involve highly super-sonic, radiative, gravity-dominated flows. In that limit, the differences owing to the algorithm are often -- though certainly not always -- minimized (see references in \S~\ref{sec:intro}). But there are important regimes with transonic flows where the numerical approach can make larger differences \citep[e.g.\ cosmological inflows \&\ outflows; see][]{vogelsberger:2011.arepo.vs.gadget.cosmo,keres:2011.arepo.gadget.disk.angmom,torrey:2011.arepo.disks}. Even in idealized test problems, we caution that simple physical differences can produce larger distinctions than the numerical method. For example, for several fluid mixing problems considered here, the ``correct'' MHD solution in the presence of an equipartition magnetic field can resemble the ``standard'' (density-entropy) SPH solution without a magnetic field, as opposed to the results from our pressure-entropy formulation or grid codes without such fields. The reason is that the real magnetic tension suppresses mixing, similar to the (purely numerical) ``surface tension'' term discussed in the text (so one might obtain a more ``realistic'' solution, but for entirely wrong reasons).

Ultimately, the numerical formulations derived here should provide the basis for a more rigorous approach to the ``flavors'' of SPH, and a means to compare the consequences of the fundamental choice of how to discretize any SPH approach. This change to the algorithm is not a panacea! Fortunately, the modified equations of motion proposed here can be trivially incorporated with many other methods that improve on other numerical aspects, for example the inviscid algorithm in \citet{cullen:2010.inviscid.sph}, the higher-order dissipation switches in \citet{price:2008.sph.contact.discontinuities,rosswog:2010.relativistic.sph} and \citet{read:2012.sph.w.dissipation.switches}, and/or the gradient error reducing integral formulation of the kernel equations in \citet{garciasenz:2012.integral.sph}. We wish to stress that -- although SPH certainly has some disadvantages which we have not attempted to address here -- poor fluid mixing in contact discontinuities is not necessarily an ``inherent'' property of SPH. This problem can be improved without requiring additional dissipation terms (and without additional computational expense) while retaining what is probably the greatest advantage of SPH algorithms, namely their excellent conservation properties. 

\vspace{-0.7cm}
\acknowledgments 
We thank Dusan Keres and Lars Hernquist for helpful discussions, and Oscar Agertz, Andrey Kravtsov, Robert Feldmann, and Nick Gnedin for contributions motivating this work. We also thank Justin Read, Volker Springel, Eliot Quataert, Claudio Dalla Vecchia, and the anonymous referee for useful comments and suggestions, and thank Andreas Bauer for sharing the implementation of the turbulent driving routine. Support for PFH was provided by NASA through Einstein Postdoctoral Fellowship Award Number PF1-120083 issued by the Chandra X-ray Observatory Center, which is operated by the Smithsonian Astrophysical Observatory for and on behalf of the NASA under contract NAS8-03060. 
\\

\vspace{-0.2cm}
\bibliography{/Users/phopkins/Documents/lars_galaxies/papers/ms}

\end{document}